\documentclass{article}

\usepackage[english]{babel}
\usepackage{upgreek}
\usepackage{cite}
\usepackage{authblk}
\usepackage{fancyhdr}
\usepackage[toc,page]{appendix}
\usepackage{epsfig,pdfpages}
\usepackage{todonotes}
\usepackage{hyperref}
\usepackage{graphicx,subfigure,relsize,xcolor}
\usepackage{amssymb,bigints,amsmath,mathtools,nicefrac}
\usepackage[infoshow,debugshow]{tabularx}
\usepackage{setspace,indentfirst}

\newcommand{\f}{\textbf{F}}
\newcommand{\C}{\textbf{C}}
\newcommand{\m}{\textbf{m}}
\newcommand{\tra}{\text{T}}
\newcommand{\sig}{\boldsymbol{\sigma}}
\newcommand{\PPi}{\boldsymbol{\Pi}}
\newcommand{\Om}{\boldsymbol{\Omega}}
\newcommand{\Lo}{\text{L}}
\newcommand{\Tv}{\text{T}}
\newcommand{\An}{\text{A}}
\newcommand{\eps}{\boldsymbol{\varepsilon}}

\newcommand{\The}{\boldsymbol{\Theta}}
\newcommand{\tpar}{\tilde{T}_{\parallel}^{\text{e}}}

\newtheorem{theorem}{Property}
\newtheorem{proof}{Proof}
\newtheorem{theo}{Method}

\DeclareMathOperator{\ti}{\text{TI}}
\DeclareMathOperator{\tr}{\text{tr}}
\DeclareMathOperator{\dev}{\text{dev}}
\DeclareMathOperator{\el}{\text{e}}
\DeclareMathOperator{\dd}{\text{d\!}}
\DeclareMathOperator{\dt}{\text{d}\tau}

\begin{document}


\title{Tensor decomposition for modified quasi-linear viscoelastic models: towards a fully nonlinear theory}

\author[1]{Valentina Balbi}
\author[2]{Tom Shearer}
\author[3]{William J Parnell}

\affil[1]{School of Mathematical and Statistical Sciences, University of Galway, University Road, Galway, Ireland}
\affil[2]{ Department of Mathematics, University of Manchester, Oxford Road, Manchester M13 9PL, United Kingdom}
\affil[3]{Department of Materials, University of Manchester, Oxford Road, Manchester M13 9PL, United Kingdom}

%
\maketitle 
\begin{abstract}
We discuss the decomposition of the tensorial relaxation function for isotropic and transversely isotropic Modified Quasi-Linear Viscoelastic models. We show how to formulate the constitutive equation by using a convenient decomposition of the relaxation tensor into scalar components and tensorial bases. We show that the bases must be \textit{symmetrically additive}, i.e they must sum up to the symmetric fourth-order identity tensor. This is a fundamental property both for isotropic and anisotropic bases that ensures the constitutive equation is consistent with the elastic limit. We provide two robust methods to obtain such bases. Furthermore, we show that, in the transversely isotropic case, the bases are naturally deformation-dependent for deformation modes that induce rotation or stretching of the fibres. Therefore, the Modified Quasi-Linear Viscoelastic framework allows to capture the non-linear phenomenon of strain-dependent relaxation, which has always been a criticised limitation of the original Quasi-Linear Viscoelastic theory. We illustrate this intrinsic non-linear feature, unique to the Modified Quasi-Linear Viscoelastic model, with two examples (uni-axial extension and perpendicular shear). 
\end{abstract}


\maketitle

\section{Introduction: linear, quasi-linear and modified quasi-linear viscoelasticity}\label{sec:intro}
According to the classical linear theory, the constitutive equation for a viscoelastic material in the integral form can be written down under the following assumptions: the material remembers the past deformation history through a fading memory, so that contributions to recent strain increments are more important than past contributions, the total stress at the current time $t$ is given by the sum of all past stress contributions
(the Boltzmann superposition principle) and the material is subjected to small deformations \cite{lakes2009viscoelastic}. Moreover, by taking the deformation history to start at $t=0$, the stress can be written as follows: 
\begin{equation}\label{sig-lin-reduced}
\sig(t)=\int_{0}^t\mathbb{G}(t-\tau):\dfrac{\dd\sig^{\el}(\tau)}{\dt}\dt,
\end{equation}
where $\sig^{\el}$ is the elastic stress tensor and $\mathbb{G}$ is the reduced relaxation tensor. $\mathbb{G}$ is fourth-order tensor whose components are the time-dependent material functions that dictate the relaxation behaviour of the material. As is the case for any fourth-order tensor, $\mathbb{G}$ can be decomposed into its components:
\begin{equation}
\mathbb{G}(t)=\sum_{n=1}^NG_n(t)\mathbb{K}_n(t),
\end{equation}
upon choosing a set of fourth-order bases $\mathbb{K}_n$, for $n=1,\dots,N$ ($N$ being the total number of independent components). The components of the reduced relaxation function satisfy the condition $G_n(0)=1$ for $n=1,\dots,N$. The functional form of the components of $\mathbb{G}$ is generally chosen with the help of experiments. A stress relaxation test can be performed where the material is suddenly deformed and then held in position for a certain time. The shape of the resulting stress curve over time will dictate the form of the relaxation function, for instance soft tissues typically display an exponentially decaying relaxation behaviour \cite{obaid2017understanding,chatelin2010fifty}. \par
Equation \eqref{sig-lin-reduced} is valid in the small deformations regime, therefore it has little application in soft tissue mechanics, where the tissues may be stretched up to twice their initial length. To overcome this limitation, Fung proposed what is now called the Quasi-Linear Viscoelastic (QLV) constitutive framework \cite{fung2013biomechanics}. The basis of Fung's idea is to extend Eq \eqref{sig-lin-reduced} to finite deformations. Mathematically, the constitutive equation is written with the respect to the second Piola-Kirchhoff stress tensor in the following form:
\begin{equation}\label{QLV-Fung}
\begin{split}
\boldsymbol{\Pi}_{\text{Fung}}(t)&=\int_0^t\mathbb{G}(t-\tau):\dfrac{\dd\boldsymbol{\Pi}^{\el}(\tau)}{\dt}\dt\\
&=\int_0^t\sum_{n=1}^N G_n(t-\tau)\mathbb{K}_n(\tau):\dfrac{\dd}{\dt}\Big(J(\tau)\f^{-1}(\tau)\textbf{T}^{\el}(\tau)\f^{-\tra}(\tau)\Big)\dt,
\end{split}
\end{equation}
where $J$ is the determinant of the deformation gradient $\f$, $\boldsymbol{\Pi}^{\el}$ is the elastic second Piola-Kirchhoff stress tensor and $\textbf{T}^{\el}$ is the elastic Cauchy stress tensor. The second Piola-Kirchhoff and the Cauchy stress tensors are linked via the Piola transformation $\boldsymbol{\Pi}=J\f^{-1}\textbf{T}\f^{-\tra}$. The QLV framework has three main advantages: (i) the constitutive equation \eqref{QLV-Fung} can account for the non-linear elastic behaviour of a material, through the tensor $\boldsymbol{\Pi}^{\el}$; (ii) the tensor $\mathbb{G}$ can be fully determined from experiments performed in the small deformation regime. This follows from the fact that the reduced relaxation tensor is inherited by the QLV model directly from the linear theory (see Eq. \eqref{sig-lin-reduced}); (iii) when it comes to modelling experimental data, a constitutive equation should be written with respect to physically measurable parameters that can be  determined straightforwardly via experiment. In principle, this can be achieved for Eq. \eqref{QLV-Fung} upon choosing an appropriate set of bases to split the reduced relaxation tensor $\mathbb{G}$ into components that are associated with specific deformation modes. \par
With this aim, the QLV theory has recently been revisited for isotropic and transversely isotropic materials in \cite{de2014nonlinear} and \cite{balbi2018modified}, respectively. In these two works, the authors proposed a modified version of the QLV theory (MQLV) where the split acts on the Cauchy stress tensor directly:
\begin{equation}\label{MQLV-Fung}
\begin{split}
\boldsymbol{\Pi}(t)&=J(t)\f^{-1}(t)\Big(\mathbb{G}(0):\textbf{T}^{\el}(t)\Big)\f^{-\tra}(t)\\
&+\int_0^t J(\tau)\f^{-1}(\tau)\Big(\mathbb{G}'(t-\tau):\textbf{T}^{\el}(\tau)\Big)\f^{-\tra}(\tau)\dt.
\end{split}
\end{equation}
This choice is motivated by the fact that the Cauchy stress has a physical meaning, representing the actual stress distribution in the material, as opposed to the second Piola-Kirchhoff stress. The problem now amounts to finding a set of bases $\mathbb{K}_n(t)$ such that the associated components $G_n(t)$ also have a physical meaning and can then be experimentally measured by common testing protocols (tensile, compression, bi-axial, shear or torsion tests).
The question of splitting isotropic and anisotropic tensors has been investigated in previous works in the context of elasticity \cite{lubarda2008elastic,spencer1984continuum}. However, as we show in this paper, these results are not applicable to the QLV framework because the splits discussed in \cite{lubarda2008elastic,spencer1984continuum} would yield a constitutive equation that does not recover the elastic limit.\par
In this paper, we show for the first time that the tensorial bases must be \textit{symmetrically additive}, i.e. they must sum up to the fourth-order symmetric identity tensor $\mathbb{S}$. Additive symmetry is a fundamental property that ensures the constitutive equation recovers the elastic limit. For the first time, we propose two different methods to derive symmetrically additive set of bases for isotropic and transversely isotropic tensors, respectively. \par 
We organise the paper as follows: in Section 2, we define the tensor algebra that we use throughout the manuscript; in Section 3, we introduce the additive symmetry property, starting by considering the isotropic case; and in Section 4, we move on to the transversely isotropic case. In Section 5, we apply the results of Sections 3 and 4 to the MQLV theory. We show that the MQLV framework naturally incorporates strain-dependent relaxation, a non-linear feature that has always been a strong limitation of the original QLV theory.
 
\section{Tensor Algebra}\label{sec:nota}
In this section, we introduce the tensor algebra that we will be using throughout the paper. All tensors are defined in $\mathbb{R}^3$. Let $\textbf{u}$ and $\textbf{v}$ be two first-order tensors ($1\!\times\!3$ vectors), $\textbf{U}$ and $\textbf{V}$ be second-order tensors ($3\!\times\!3$ matrices) with $\textbf{U}$ being symmetric ($\textbf{U}=\textbf{U}^{\tra}$), and $\mathbb{U}$ and $\mathbb{V}$ be $3\!\times\!3\!\times\!3\!\times\!3$ fourth-order tensors. We define the identity matrix $\textbf{I}$ with components $\text{I}_{ab}=\delta_{ab}$, where $\delta_{ab}=1$ if $a=b$ and is zero otherwise and $a$ and $b$ take the values $\{1,2,3\}$. We work with the following fourth-order identity tensors: 
\begin{equation}
\begin{split}
&\overline{\mathbb{I}}=\textbf{I}\overline\otimes\textbf{I} \hspace{2.35cm} \text{or} \hspace{1cm}\overline{\mathbb{I}}_{abcd}=\delta_{ad}\delta_{bc},\\
&\underline{\mathbb{I}}=\textbf{I}\underline{\otimes}\textbf{I} \hspace{2.35cm} \text{or} \hspace{1cm}\underline{\mathbb{I}}_{abcd}=\delta_{ac}\delta_{bd},
\end{split}
\end{equation}
and the symmetric fourth-order tensor:
\begin{equation}\label{esse}
\mathbb{S}=\dfrac{1}{2}(\textbf{I}\overline{\otimes}\textbf{I}+\textbf{I}\underline{\otimes}\textbf{I}) \hspace{1cm} \text{or} \hspace{1cm}\mathbb{S}_{abcd}=\dfrac{1}{2}(\delta_{ad}\delta_{bc}+\delta_{ac}\delta_{bd}).
\end{equation}
The tensor $\mathbb{S}$ has the following major and minor symmetries:
\begin{equation}\label{symmS}
\mathbb{S}_{abcd}=\mathbb{S}_{cdab}\quad \text{and}\quad \mathbb{S}_{abcd}=\mathbb{S}_{abdc}=\mathbb{S}_{bacd},
\end{equation}
respectively, whereas the tensors $\overline{\mathbb{I}}$ and $\underline{\mathbb{I}}$ only have the major symmetries:
\begin{equation}\label{symmIs}
\begin{split}
&\overline{\mathbb{I}}_{abcd}=\overline{\mathbb{I}}_{cdab},\\
&\underline{\mathbb{I}}_{abcd}=\underline{\mathbb{I}}_{cdab}.
\end{split}
\end{equation}
For any second-order symmetric tensor $\textbf{U}$ the following rules apply:
\begin{align}
&\mathbb{S}:\textbf{U}=\dfrac{1}{2}(\textbf{U}+\textbf{U}^{\tra})=\textbf{U}=\underline{\mathbb{I}}:\textbf{U},\\\label{sym}
&\mathbb{S}:\textbf{U}=\dfrac{1}{2}(\textbf{U}+\textbf{U}^{\tra})=\textbf{U}^{\tra}=\overline{\mathbb{I}}:\textbf{U}.
\end{align}
Moreover, we use the following tensor rules: 
\begin{equation}
\begin{split}
&(\textbf{u}\otimes\textbf{v})_{ab}=\text{u}_a\text{v}_b,\hspace{1.65cm}\textbf{U}:\textbf{V}=\text{U}_{ab}\text{V}_{ab},\hspace{1.3cm}
(\mathbb{U}:\textbf{V})_{ab}=\mathbb{U}_{abcd}\text{V}_{cd},\\
&(\textbf{U}\overline\otimes\textbf{V})_{abcd}=\text{U}_{ad}\text{V}_{bc},
\hspace{1cm}
(\textbf{U}\underline\otimes\textbf{V})_{abcd}=\text{U}_{ac}\text{V}_{bd},\hspace{0.5cm}
(\textbf{U}\otimes\textbf{V})_{abcd}=\text{U}_{ab}\text{V}_{cd},\\
&(\overline{\mathbb{I}}:\textbf{V})_{ab}=\text{V}_{ba},\hspace{2cm}
(\underline{\mathbb{I}}:\textbf{V})_{ab}=\text{V}_{ab},
\hspace{1.45cm}
(\mathbb{I}:\textbf{V})_{ab}=\dfrac{1}{2}(\text{V}_{ab}+\text{V}_{ba}),\\
&(\mathbb{U}:\mathbb{V})_{abcd}=\mathbb{U}_{abxy}\mathbb{V}_{xycd},
\end{split}
\end{equation}
where repeated indices imply summation.

\section{Fourth-order bases for isotropic materials}\label{sec:iso}
We start by considering the isotropic case. From the theory of elasticity, we know that the elasticity tensor for an isotropic material has 2 independent components. The material properties of a linearly elastic material are uniquely identified by any of the following pairs of parameters: the first Lam\'e constant $\lambda$ and the shear modulus $\mu$, the bulk modulus $\kappa$ and $\mu$, the Young's modulus $E$ and the Poisson's ratio $\nu$. Each pair is associated to a different set of bases. For instance, the set of bases:
\begin{equation}\label{set1-iso}
\begin{split}
&\mathbb{I}_1=\dfrac{1}{3}\textbf{I}\otimes\textbf{I}\hspace{3cm}\text{or}\hspace{1cm}\mathbb{I}_{1abcd}=\dfrac{1}{3}\delta_{ab}\delta_{cd},\\
&\mathbb{I}_2=\dfrac{1}{2}(\textbf{I}\overline{\otimes}\textbf{I}+\textbf{I}\underline{\otimes}\textbf{I}) -\dfrac{1}{3}\textbf{I}\otimes\textbf{I}\hspace{0.7cm}\text{or}\hspace{1cm}\mathbb{I}_{2abcd}=
\dfrac{1}{2}(\delta_{ad}\delta_{bc}+\delta_{ac}\delta_{bd})-\dfrac{1}{3}\delta_{ab}\delta_{cd},
\end{split}
\end{equation}
yields the following constitutive equation for a linearly elastic isotropic material:
\begin{equation}\label{sig-el-1}
\sig^e=\mathbb{C}:\eps=\sum_{n=1}^2C_n\,\mathbb{I}_n:\eps=\kappa\tr[\eps]\textbf{I}+2\mu\dev[\eps],
\end{equation}
where $C_1=3\kappa$ and $C_2=2\mu$ and $\dev[\eps]=\eps-\dfrac{1}{3}\tr[\eps]\textbf{I}$.\par
The bases $\{\mathbb{I}_1,\mathbb{I}_2\}$ have the following important properties: idempotence, orthogonality and additive symmetry.
\begin{theorem} \textbf{Idempotence} under double contraction	:
\begin{equation}\label{prop-idem}
\mathbb{I}_1:\mathbb{I}_1=\mathbb{I}_1,\qquad\mathbb{I}_2:\mathbb{I}_2=\mathbb{I}_2.
\end{equation}
\end{theorem}
\begin{proof}
$\forall a, b, c, d$, the double contraction of $\mathbb{I}_1$ with itself gives:
\begin{equation*}
\begin{split}
(\mathbb{I}_1:\mathbb{I}_1)_{abcd}&=\mathbb{I}_{1abxy}\mathbb{I}_{1xycd}=\sum_{x=1}^3\sum_{y=1}^3\dfrac{1}{3}\delta_{ab}\delta_{xy}\dfrac{1}{3}\delta_{xy}\delta_{cd}=\dfrac{1}{9}\delta_{ab}\delta_{cd}\sum_{x=1}^3\sum_{y=1}^3\delta_{xy}\delta_{xy}\\
&=\dfrac{1}{3}\delta_{ab}\delta_{cd}=\mathbb{I}_{1abcd}.
\end{split}
\end{equation*}
Similarly, the double contraction of $\mathbb{I}_2$ with itself yields:
\begin{equation*}
\begin{split}
(\mathbb{I}_2:\mathbb{I}_2)_{abcd}&=\sum_{x=1}^3\sum_{y=1}^3\Big(\dfrac{1}{2}\delta_{ay}\delta_{bx}+\dfrac{1}{2}\delta_{ax}\delta_{by}-\dfrac{1}{3}\delta_{ab}\delta_{xy}\Big)\Big(\dfrac{1}{2}\delta_{xd}\delta_{yc}+\dfrac{1}{2}\delta_{xc}\delta_{yd}-\dfrac{1}{3}\delta_{xy}\delta_{cd}\Big)\\
&=\sum_{x=1}^3\sum_{y=1}^3\Big(\dfrac{1}{4}\delta_{ay}\delta_{yc}\delta_{bx}\delta_{xd}+\dfrac{1}{4}\delta_{ay}\delta_{yd}\delta_{bx}\delta_{xc}-\dfrac{1}{6}\delta_{ay}\delta_{xy}\delta_{bx}\delta_{cd}\\
&\hspace{1.3cm}+\dfrac{1}{4}\delta_{ax}\delta_{xd}\delta_{by}\delta_{yc}+\dfrac{1}{4}\delta_{ax}\delta_{xc}\delta_{by}\delta_{yd}-\dfrac{1}{6}\delta_{ax}\delta_{xy}\delta_{by}\delta_{cd}\\
&\hspace{1.3cm}-\dfrac{1}{6}\delta_{ab}\delta_{xy}\delta_{xd}\delta_{yc}-\dfrac{1}{6}\delta_{ab}\delta_{xy}\delta_{xc}\delta_{yd}+\dfrac{1}{9}\delta_{ab}\delta_{xy}\delta_{xy}\delta_{cd}\Big)\\
&=\dfrac{1}{2}(\delta_{ac}\delta_{bd}+\delta_{ad}\delta_{bc})-\dfrac{1}{3}\delta_{ab}\delta_{cd}=\mathbb{I}_{2abcd}.
\end{split}
\end{equation*}
\end{proof}
\begin{theorem} \textbf{Orthogonality}:
\begin{equation}\label{prop-ortho}
\mathbb{I}_1:\mathbb{I}_2=\mathbb{O},\qquad\mathbb{I}_2:\mathbb{I}_1=\mathbb{O},
\end{equation}
where $\mathbb{O}$ is the fourth-order null tensor, with components $\mathbb{O}_{abcd}=0$, $\forall a,b,c,d$.
\end{theorem}
\begin{proof}
Similarly, as in the previous proof, we calculate the contraction between the two bases in components:
\begin{equation*}
\begin{split}
(\mathbb{I}_1:\mathbb{I}_2)_{abcd}&=\sum_{x=1}^3\sum_{y=1}^3\Big(\dfrac{1}{3}\delta_{ab}\delta_{xy}\Big)\Big(\dfrac{1}{2}\delta_{xd}\delta_{yc}+\dfrac{1}{2}\delta_{xc}\delta_{yd}-\dfrac{1}{3}\delta_{xy}\delta_{cd}\Big)\\
&=\sum_{x=1}^3\sum_{y=1}^3\Big(\dfrac{1}{6}\delta_{ab}\delta_{xy}\delta_{xd}\delta_{yc}+\dfrac{1}{6}\delta_{ab}\delta_{xy}\delta_{xc}\delta_{yd}
-\dfrac{1}{9}\delta_{ab}\delta_{xy}\delta_{xy}\delta_{cd}\Big)\\
&=\dfrac{2}{6}\delta_{ab}\delta_{cd}-\dfrac{1}{3}\delta_{ab}\delta_{cd}=0,
\end{split}
\end{equation*}
and
\begin{equation*}
\begin{split}
(\mathbb{I}_2:\mathbb{I}_1)_{abcd}&=\sum_{x=1}^3\sum_{y=1}^3\Big(\dfrac{1}{2}\delta_{ay}\delta_{bx}+\dfrac{1}{2}\delta_{ax}\delta_{by}-\dfrac{1}{3}\delta_{ab}\delta_{xy}\Big)\Big(\dfrac{1}{3}\delta_{xy}\delta_{cd}\Big)\\
&=\sum_{x=1}^3\sum_{y=1}^3\Big(\dfrac{1}{6}\delta_{ay}\delta_{xy}\delta_{bx}\delta_{cd}+\dfrac{1}{6}\delta_{ax}\delta_{xy}\delta_{by}\delta_{cd}
-\dfrac{1}{9}\delta_{ab}\delta_{xy}\delta_{xy}\delta_{cd}\Big)\\
&=\dfrac{2}{6}\delta_{ab}\delta_{cd}-\dfrac{1}{3}\delta_{ab}\delta_{cd}=0.
\end{split}
\end{equation*}
\end{proof}
\begin{theorem}\label{prop-3}
\textbf{Additive symmetry}:
\begin{equation}\label{prop-symm-add}
\mathbb{I}_1+\mathbb{I}_2=\mathbb{S},
\end{equation}
where $\mathbb{S}$ is the symmetric fourth-order identity tensor in Eq.\eqref{esse}.
\end{theorem}
\begin{proof}
Again we can easily show that $\forall a, b, c, d$:
\begin{equation*}
\begin{split}
(\mathbb{I}_1+\mathbb{I}_2)_{abcd}&=\mathbb{I}_{1abcd}+\mathbb{I}_{2abcd}=\dfrac{1}{3}\delta_{ab}\delta_{cd}+\Big(\dfrac{1}{2}\delta_{ad}\delta_{bc}+\dfrac{1}{2}\delta_{ac}\delta_{bd}-\dfrac{1}{3}\delta_{ab}\delta_{cd}\Big)\\
&=\dfrac{1}{2}\Big(\delta_{ad}\delta_{bc}+\delta_{ac}\delta_{bd}\Big)=\mathbb{S}_{abcd}.
\end{split}
\end{equation*}
\end{proof}

For symmetric tensors $\textbf{U}=\textbf{U}^{\tra}$, as shown in Eq. \eqref{sym}, the tensor $\mathbb{S}$ acts as the fourth-order identity tensor $\underline{\mathbb{I}}$. Therefore:
\begin{equation}\label{prop-add}
\underline{\mathbb{I}}:\textbf{U}=\mathbb{S}:\textbf{U}=(\mathbb{I}_1+\mathbb{I}_2):\textbf{U}=\dfrac{1}{3}\tr[\textbf{U}]\textbf{I}+\big(\textbf{U}-\dfrac{1}{3}\tr[\textbf{U}]\textbf{I}\big)=\textbf{U}.
\end{equation}
Eq. \eqref{prop-add} shows that the set of symmetrically additive bases $\{\mathbb{I}_1,\mathbb{I}_2\}$ can be used to split a symmetric second-order tensor into the sum of two terms. Although this is a well-known result, it plays a crucial role in 
the visco-elastic setting, as we show later in the manuscript. \\

Let us now consider the set of bases:
\begin{equation}\label{set2-iso}
\mathbb{\hat{I}}_1=\textbf{I}\otimes\textbf{I}=3\mathbb{I}_1
\hspace{1cm}\text{and}\hspace{1cm}\mathbb{\hat{I}}_2=\dfrac{1}{2}(\textbf{I}\overline{\otimes}\textbf{I}+\textbf{I}\underline{\otimes}\textbf{I})=\mathbb{I}_1+\mathbb{I}_2.
\end{equation}
They are linked to the bases in \eqref{set1-iso} by the following connections: $\mathbb{\hat{I}}_1=3\mathbb{I}_1$ and $\mathbb{\hat{I}}_2=\mathbb{I}_2+\mathbb{I}_1$. The set of bases  in \eqref{set2-iso} allows us to write the constitutive equation for a linear elastic material as follows:
\begin{equation}\label{sig-el-2}
\sig^e=\mathbb{C}:\eps=\sum_{n=1}^N\hat{C}_n\,\mathbb{\hat{I}}_n:\eps=\lambda\tr[\eps]\textbf{I}+2\mu\eps,
\end{equation}
where $\hat{C}_1=\lambda$ and $\hat{C}_2=2\mu$. By comparing Eqs. \eqref{sig-el-1} and \eqref{sig-el-2} we obtain the following links between the components $C_n$ ($n=1,2$) associated with the set of bases in Eq. \eqref{set1-iso} and the components $\hat{C}_n$ associated with the bases in Eq. \eqref{set2-iso}, i.e.: 
\begin{equation}\label{conv-el-par}
\hat{C}_1=\dfrac{C_1-C_2}{3}\hspace{1cm}\text{and}\hspace{1cm}\hat{C}_2=C_2.
\end{equation}
From Eq. \eqref{conv-el-par}, we can recover the well-known relation between the first Lam\'e constant and the bulk modulus, $\lambda=\kappa-\nicefrac{2}{3}\mu$. We can easily verify that, although the two constitutive equations in \eqref{sig-el-1} and \eqref{sig-el-2} are equivalent, the bases in Eqs. \eqref{set1-iso} and \eqref{set2-iso} do not share the same properties. Indeed, the set in Eq. \eqref{set2-iso} has none of the three properties. In particular, the bases $\{\hat{\mathbb{I}}_1,\hat{\mathbb{I}}_2\}$ do not have the additive symmetry property:
\begin{equation}\label{set2-nogood}
(\hat{\mathbb{I}}_1+\hat{\mathbb{I}}_2):\textbf{U}=\tr[\textbf{U}]\textbf{I}+\textbf{U}\neq\textbf{U}.
\end{equation}

\subsection{Linear viscoelasticity}
Let us now consider the viscoelastic setting. In the same fashion as for the elastic setting, we want to find some set of bases for the tensor $\mathbb{G}$ that allows us to rewrite the constitutive equation in \eqref{sig-lin-reduced} with respect to different relaxation functions. In particular, we would like the relaxation functions to be associated with deformation modes that can be performed experimentally. The constitutive form \eqref{sig-lin-reduced} is written with respect to the elastic stress and the reduced relaxation tensor and is a convenient form to later make the link with the MQLV theory.  \par 
Integration by parts of Eq. \eqref{sig-lin-reduced} gives:
\begin{equation}
\sigma(t)=\underbrace{\mathbb{G}(0):\sig^{\el}(t)}_{\text{elastic stress}}+\int_0^t\mathbb{G}'(t-\tau):\sig^{\el}(\tau)\dt.
\end{equation}
Since $\mathbb{G}$ is the reduced relaxation tensor, its components satisfy the condition $G_n(0)=1,\forall n$. Now let $\{\mathbb{K}_1,\mathbb{K}_2\}$ be the set of bases associated with the components $\{G_1,G_2\}$, then the term outside the integral must give the elastic stress $\sig^{\el}(t)$: 
\begin{equation}
\sum_{n=0}^2G_n(0)\mathbb{K}_n:\sig^{\el}(t)=\sum_{n=0}^2\mathbb{K}_n:\sig^{\el}(t)\equiv\sig^{\el}(t).
\end{equation}
Therefore, the bases $\mathbb{K}_n$ must posses the property of  \textbf{additive symmetry}. This property is crucial in order for the constitutive equation to recover the elastic limit. \par
We can now show how to write the constitutive equation for a linear viscoelastic material in different forms, each with different relaxation functions. Since the bases $\{\mathbb{I}_1,\mathbb{I}_2\}$ in Eq. \eqref{set1-iso} are symmetrically additive, we can use them to split the tensor $\mathbb{G}$, as follows:
\begin{equation}\label{sig-ve-1}
\begin{split}
\sig(t)&=\int_0^t\mathbb{G}(t-\tau):\dfrac{\dd \sig^{\el}(\tau)}{\dt}\dt=\int_0^tG_1(t-\tau)\mathbb{I}_1:\dfrac{\dd \sig^{\el}(\tau)}{\dt}\dt+\int_0^tG_2(t-\tau)\mathbb{I}_2:\dfrac{\dd \sig^{\el}(\tau)}{\dt}\dt\\
&=\int_0^tG_1(t-\tau)\mathbb{I}_1:\dfrac{\dd \Big(\sum\limits_{n=1}^2 C_n\mathbb{I}_n:\eps(\tau)\Big)}{\dt}\dt+\int_0^tG_2(t-\tau)\mathbb{I}_2:\dfrac{\dd \Big(\sum\limits_{n=1}^2 C_n\mathbb{I}_n:\eps(\tau)\Big)}{\dt}\dt\\
&=\int_0^tG_1(t-\tau)C_1\mathbb{I}_1:\dfrac{\dd\eps(\tau)}{\dt}\dt+\int_0^tG_2(t-\tau)C_2\mathbb{I}_2:\dfrac{\dd\eps(\tau)}{\dt}\dt\\
&=\int_0^tG_1(t-\tau)\kappa\dfrac{\dd\,\tr[\eps(\tau)]\textbf{I}}{\dt}\dt+\int_0^tG_2(t-\tau)(2\mu)\dfrac{\dd\,\dev[\eps(\tau)]}{\dt}\dt\\
&=\int_0^t\kappa(t-\tau)\dfrac{\dd\,\tr[\eps(\tau)]\textbf{I}}{\dt}\dt+2\int_0^t\mu(t-\tau)\dfrac{\dd\,\dev[\eps(\tau)]}{\dt}\dt.
\end{split}
\end{equation}
In Eq. \eqref{sig-ve-1}, we used Eq. \eqref{sig-el-1} to write the elastic stress $\sig^{\el}$ in terms of the bulk and shear moduli, and the idempotent and the orthogonal properties of the bases $\{\mathbb{I}_1,\mathbb{I}_2\}$, which are given in Eqs. \eqref{prop-idem} and \eqref{prop-ortho}, respectively. Moreover, we have now a link between the components of the reduced relaxation tensor, i.e. the non-dimensional relaxation functions $G_1(t)$ and $G_2(t)$, and the relaxation functions associated with the bulk and the shear modulus, $\kappa(t)$ and $\mu(t)$, respectively:
\begin{equation}\label{relax-func-iso}
G_1(t)=\dfrac{\kappa(t)}{\kappa}\qquad\text{and}\qquad G_2(t)=\dfrac{\mu(t)}{\mu}.
\end{equation}
The advantage of Eq. \eqref{sig-ve-1} is that the constitutive equation is written with respect to two relaxation functions that have a direct physical interpretation and can be determined experimentally by performing two well-established mechanical tests: a simple shear (or torsion) test to determine $\mu(t)$ and a hydrostatic compression to determine $\kappa(t)$.\par
Similarly, we can obtain an equivalent form for $\sig$ by using the set of bases in Eq. \eqref{set1-iso} to split the tensor $\mathbb{G}$ and the set of bases in Eq. \eqref{set2-iso} to write $\sig^{\el}$, as follows:
\begin{equation}\label{sig-ve-2}
\begin{split}
\sig(t)&=\int_0^tG_1(t-\tau)\mathbb{I}_1:\dfrac{\dd \Big(\sum\limits_{n=1}^2 \hat{C}_n\hat{\mathbb{I}}_n:\eps(\tau)\Big)}{\dt}\dt+\int_0^tG_2(t-\tau)\mathbb{I}_2:\dfrac{\dd \Big(\sum\limits_{n=1}^2 \hat{C}_n\hat{\mathbb{I}}_n:\eps(\tau)\Big)}{\dt}\dt\\
&=\int_0^tG_1(t-\tau)(\hat{C}_1+\dfrac{1}{3}\hat{C}_2)\hat{\mathbb{I}}_1:\dfrac{\dd\eps(\tau)}{\dt}\dt+\int_0^tG_2(t-\tau)\hat{C}_2\Big(\mathbb{\hat{I}}_2-\dfrac{1}{3}\mathbb{\hat{I}}_1\Big):\dfrac{\dd\eps(\tau)}{\dt}\dt\\
&=\int_0^t\Big(G_1(t-\tau)\kappa-\dfrac{2}{3}G_2(t-\tau)\mu\Big)\dfrac{\dd\,\tr[\eps(\tau)]\textbf{I}}{\dt}\dt+\int_0^tG_2(t-\tau)(2\mu)\dfrac{\dd\,\eps(\tau)}{\dt}\dt\\
&=\int_0^t\lambda(t-\tau)\dfrac{\dd\,\tr[\eps(\tau)]\textbf{I}}{\dt}\dt+2\int_0^t\mu(t-\tau)\dfrac{\dd\,\eps(\tau)}{\dt}\dt,
\end{split}
\end{equation}
where we used the links between $\hat{C}_n$ and $C_n$, $n=\{1,2\}$ in Eq. \eqref{conv-el-par}, Eq. \eqref{relax-func-iso} and the results: 
\begin{align}
&\mathbb{I}_1:\hat{\mathbb{I}}_1=\hat{\mathbb{I}}_1:\mathbb{I}_1=\hat{\mathbb{I}}_1,\qquad\mathbb{I}_1:\mathbb{\hat{I}}_2=\mathbb{\hat{I}}_2:\mathbb{I}_1=\nicefrac{1}{3}\,\hat{\mathbb{I}}_1,\\
&\mathbb{I}_2:\hat{\mathbb{I}}_1=\hat{\mathbb{I}}_1:\mathbb{I}_2=\mathbb{O},\qquad\mathbb{I}_2:\mathbb{\hat{I}}_2=\mathbb{\hat{I}}_2:\mathbb{I}_2=\hat{\mathbb{I}}_2-\nicefrac{1}{3}\,\hat{\mathbb{I}}_1.
\end{align}
By comparing Eqs. \eqref{sig-ve-1} with \eqref{sig-ve-2}, we obtain the link between the relaxation function $\kappa(t)$ associated with the bulk modulus and the relaxation function $\lambda(t)$ associated to the first Lam\'e parameter: $\lambda(t)=\kappa(t)-\dfrac{2}{3}\mu(t), \forall t$.\par
Finally, we note that, since the bases in Eq. \eqref{set2-iso} are not symmetrically additive, we cannot use them to split the tensor $\mathbb{G}$. Indeed, they cannot be used to derive either of the two constitutive equations in \eqref{sig-ve-1} and \eqref{sig-ve-2}. 

\subsection{How to obtain a symmetrically additive set of bases from a non-symmetrically additive set}
Now, the question of whether it is possible to derive a set of symmetrically additive bases from a set of non-symmetrically additive bases arises naturally from the analysis carried out in the previous sections. We show here a simple method that works for the isotropic case and in the next section we will extend it to give a more general method that can be applied to anisotropic fourth-order tensors. \par
The problem can be stated as follows. Given a set of non-symmetrically additive isotropic bases $\{\hat{\mathbb{K}}_1,\hat{\mathbb{K}}_2\}$, the requirement is to find a set of symmetrically additive isotropic bases $\{\mathbb{K}_1,\mathbb{K}_2\}$, such that 
$\mathbb{K}_1+\mathbb{K}_2=\mathbb{S}$.

\begin{theo}\label{M1}
We first write the symmetricall aditive bases as a linear combination of  the non-symmetrically additive bases $\{\hat{\mathbb{K}}_1,\hat{\mathbb{K}}_2\}$ as follows:
\begin{equation}\label{ruleISO}
\left\{\begin{array}{l}
\mathbb{K}_1=a_{11}\hat{\mathbb{K}}_1+a_{12}\hat{\mathbb{K}}_2,\\
\mathbb{K}_2=a_{21}\hat{\mathbb{K}}_1+a_{22}\hat{\mathbb{K}}_2,
\end{array}\right.
\end{equation}
Then we impose the additive symmetry property by requiring that:
\begin{equation}\label{MainEqISO}
\underbrace{(a_{11}+a_{21})\hat{\mathbb{K}}_1}_{\mathbb{K}_1}+\underbrace{(a_{12}+a_{22})\hat{\mathbb{K}}_2}_{\mathbb{K}_2}-\mathbb{S}=\mathbb{O},
\end{equation}
and we solve Eq. \eqref{MainEqISO} to obtain the coefficients $a_{ij}$.
\end{theo}
To illustrate Method \ref{M1}, we derive a symmetrically additive set of bases from the bases $\{\hat{\mathbb{I}}_1,\hat{\mathbb{I}}_2\}$ in Eq. \eqref{set2-iso}.
The tensors $\hat{\mathbb{I}}_1,\hat{\mathbb{I}}_2$ and $\mathbb{S}$ all share the symmetries in Eqs. \eqref{symmS}, therefore Eq. \eqref{MainEqISO} reduces to the following 2 equations:
\begin{equation}\label{sys-addsymm-Iso}
\left\{\begin{array}{l}
a_{11} + a_{21}=0,\\
a_{12} + a_{22}=1.
\end{array}\right.
\end{equation}
Since the number of unknowns is greater than the number of equations, there are infinite solutions to the system in Eq. \eqref{sys-addsymm-Iso}, namely $a_{21} =-a_{11}$ and $ a_{22} = 1 - a_{12}$. However, we can use Properties \ref{prop-idem} and \ref{prop-ortho} to restrict further the number of solutions and require that:
\begin{align}
\mathbb{K}_1:\mathbb{K}_1=\mathbb{K}_1 \quad\Leftrightarrow\quad a_{11}^2\hat{\mathbb{I}}_1:\hat{\mathbb{I}}_1+2a_{11}a_{12}\hat{\mathbb{I}}_1:\hat{\mathbb{I}}_2+a_{12}^2\hat{\mathbb{I}}_2:\hat{\mathbb{I}}_2=a_{11}\hat{\mathbb{I}}_1+a_{12}\hat{\mathbb{I}}_2,\label{idem1}\\ 
\mathbb{K}_2:\mathbb{K}_2=\mathbb{K}_2 \quad\Leftrightarrow\quad a_{21}^2\hat{\mathbb{I}}_1:\hat{\mathbb{I}}_1+2a_{21}a_{22}\hat{\mathbb{I}}_1:\hat{\mathbb{I}}_2+a_{22}^2\hat{\mathbb{I}}_2:\hat{\mathbb{I}}_2=a_{21}\hat{\mathbb{I}}_1+a_{22}\hat{\mathbb{I}}_2, \label{idem2}
\end{align}
and
\begin{equation}\label{ortho-eq}
\mathbb{K}_1:\mathbb{K}_2=\mathbb{O}\quad\Leftrightarrow\quad a_{11}a_{21}\hat{\mathbb{I}}_1:\hat{\mathbb{I}}_1+(a_{11}a_{22}+a_{12}a_{21})\hat{\mathbb{I}}_1:\hat{\mathbb{I}}_2+a_{12}a_{22}\hat{\mathbb{I}}_2:\hat{\mathbb{I}}_2=\mathbb{O},
\end{equation}
respectively. Each of Eqs. \eqref{idem1} and \eqref{idem2} reduce to a non-linear system of 2 equations in the following form:
\begin{equation}\label{sys-idem-iso}
\left\{\begin{array}{l}
a_{11}(3 a_{11} + 2 a_{12}-1)=0,\\
a_{12}(a_{12}-1)=0,
\end{array}\right.\qquad\text{and}\qquad
\left\{\begin{array}{l}
a_{21}(3 a_{21} + 2 a_{22}-1)=0,\\
a_{22}(a_{22}-1)=0,
\end{array}\right.\hfill
\end{equation}
respectively.
Eq. \eqref{ortho-eq} reduces to the following non-linear system:
\begin{equation}\label{sys-ortho-iso}
\left\{\begin{array}{l}
a_{12} a_{21} + a_{11} (3 a_{21} + a_{22})=0\\
a_{12} a_{22}=0.
\end{array}\right.
\end{equation}
By combining \eqref{sys-addsymm-Iso} with the first system of \eqref{sys-idem-iso}, we obtain the second system of \eqref{sys-idem-iso} and vice-versa by combining \eqref{sys-addsymm-Iso} with the second system of \eqref{sys-idem-iso} we obtain the first system in \eqref{sys-idem-iso}. Therefore, under the additive symmetry property, if one of the bases is idempotent, then the other is as well. Moreover, by combining \eqref{sys-addsymm-Iso}  with \eqref{sys-ortho-iso}, we obtain the second system in \eqref{sys-idem-iso}. Therefore, if the bases are orthogonal, then they are also idempotent and vice-versa. The systems in \eqref{sys-addsymm-Iso}, \eqref{sys-idem-iso} and \eqref{sys-ortho-iso} have the following 4 sets of solutions:
\begin{align}
\textbf{Sol-1:}\quad&a_{11} = a_{12}=a_{21}= 0,\quad  a_{22}= 1,\label{triv1}\\
\textbf{Sol-2:}\quad&a_{11} =a_{21}= a_{22}= 0, \quad a_{12}= 1,\label{triv2}\\
\textbf{Sol-3:}\quad&a_{11} = -\dfrac{1}{3}, \quad a_{12}= 1, \quad a_{21}= \dfrac{1}{3},\quad  a_{22}= 0,\label{sol3}\\
\textbf{Sol-4:}\quad&a_{11} = \dfrac{1}{3}, \quad a_{12}= 0, \quad a_{21}= -\dfrac{1}{3},\quad  a_{22}= 1.\label{sol4}
\end{align}
Sol-1 and Sol-2 can be discarded as they result in one of the two bases being zero. Sol-3 and Sol-4 are equivalent and give the symmetrically additive set in Eq. \eqref{set1-iso} which is also idempotent and orthogonal. \par 
In conclusion, in this section we have proposed a simple method for deriving a set of symmetrically additive bases from a non-symmetrically additive set for isotropic tensors. The method requires us to express the unknown symmetrically additive set as a linear combination of the known non-symmetrically additive bases and calculate the linear coefficients by imposing Property \ref{prop-symm-add}. As expected, there are infinite combinations of bases that satisfy the symmetrically additive property; however, there is only one set (given by Sol-3 and Sol-4 in Eqs. \eqref{sol3} and \eqref{sol4}) that is both idempotent (Property \ref{prop-idem}) and orthogonal (Property \ref{prop-ortho}). \par
We now investigate how this procedure can be extended to the anisotropic case. In particular, we show that Method 1 is not sufficient in this case and an alternative route is required to construct a set of basis that is symmetrically additive.

\section{Fourth-order bases for transversely isotropic materials}\label{sec:TI}
Let us now focus on transversely isotropic (TI) materials. Such anisotropic materials typically either contain a family of fibres or a layered structure that produces mechanical behaviour along a single axis that is distinct from that in the transverse directions. We identify this preferred direction with the unit vector $\m$. The elasticity tensor $\mathbb{C}^{\ti}$ for a TI material has 6 components, of which 5 are independent due to its symmetries.\par
As in the previous section, we start by writing the constitutive equation for a linear elastic material, but now in the case of transverse isotropy.
The following set of bases was originally proposed by Spencer \cite{spencer1984continuum}:
\begin{equation}\label{set-TI-1}
\begin{split}
&\hat{\mathbb{J}}_1=\textbf{I}\otimes\textbf{I},\quad\hat{\mathbb{J}}_2=\textbf{I}\otimes(\m\otimes\m),\quad\hat{\mathbb{J}}_3=(\m\otimes\m)\otimes\textbf{I},\quad\hat{\mathbb{J}}_4=(\m\otimes\m)\otimes(\m\otimes\m),\\
&\hat{\mathbb{J}}_5=\dfrac{1}{2}(\overline{\mathbb{I}}+\underline{\mathbb{I}})-\dfrac{1}{2}(\textbf{I}\overline{\otimes}(\m\otimes\m)+\textbf{I}\underline{\otimes}(\m\otimes\m)+(\m\otimes\m)\overline{\otimes}\textbf{I}+(\m\otimes\m)\underline{\otimes}\textbf{I}),\\
&\hat{\mathbb{J}}_6=\dfrac{1}{2}(\overline{\mathbb{I}}+\underline{\mathbb{I}})-\hat{\mathbb{J}}_5,
\end{split}
\end{equation}
and yields to the well-known form:
\begin{equation}\label{sig-el-TI-1}
\begin{split}
\sig^{\el}=\mathbb{C}^{\ti}:\eps=\sum_{n=1}^6\hat{C}^{\ti}_n\,\hat{\mathbb{J}}_n:\eps&=\big(\lambda\tr[\eps]+\alpha\,\m\!\cdot\!\eps\m\big)\textbf{I}+\big(\alpha\tr[\eps]+\beta\,\m\!\cdot\!\eps\m\big)\m\otimes\m\\
&+2\mu_t\big(\eps-\eps\m\otimes\m-\m\otimes\eps\m\big)+2\mu_l(\eps\m\otimes\m+\m\otimes\eps\m),
\end{split}
\end{equation}
where $\mu_t$ and $\mu_l$ are the shear moduli associated with shear in the plane of isotropy and the planes containing the axis of anisotropy, respectively, $\alpha$ and $\beta$ are two parameters related to the Young's moduli of the isotropic matrix and the fibres, respectively.
Another set of TI bases that is commonly used in the continuum mechanics community is given by the Hill bases \cite{parnell2016eshelby,mahnken2005anisotropy}:
\begin{equation}\label{set-TI-Hill}
\begin{split}
\hat{\mathbb{H}}_1 &= \frac{1}{2}\The\otimes\The, \quad
\hat{\mathbb{H}}_2 = \The\otimes(\m\otimes\m),\quad
\hat{\mathbb{H}}_3 =(\m\otimes\m)\otimes\The,\quad
\hat{\mathbb{H}}_4 = \m\otimes\m\otimes\m\otimes\m\\
\hat{\mathbb{H}}_5 &= \frac{1}{2}(\The\overline{\otimes}\The+\The\underline{\otimes}\The-\The\otimes\The),\\
\hat{\mathbb{H}}_6 &= \frac{1}{2}(\The\overline{\otimes}(\m\otimes\m)+\The\underline{\otimes}(\m\otimes\m)+(\m\otimes\m)\overline{\otimes}\The+(\m\otimes\m)\underline{\otimes}\The),
\end{split}
\end{equation}
where $\The=\textbf{I}-\m\otimes\m$.
We can easily verify that both the bases in Eq. \eqref{set-TI-1} and in Eq. \eqref{set-TI-Hill} are not symmetrically additive. As opposed to the isotropic case, a symmetrically additive set of bases for transversely isotropic tensors has not yet been proposed in the literature. Our aim here is to construct such a set together with a general method for deriving a symmetrically additive set of bases for TI tensors. Method \ref{M1} is easy to implement in the isotropic case, where the number of unknowns, i.e. the 4 coefficients in Eq. \eqref{ruleISO}, is small. However, in the TI case, Method 1 would lead us to state the problem as follows.
Given a set of non-symmetricallly additive TI bases $\{\hat{\mathbb{K}}_1,\dots,\hat{\mathbb{K}}_6\}$, we require to find a set of symmetrically additive TI bases $\{\mathbb{K}_1,\dots,\mathbb{K}_6\}$ as a linear combination of $\{\hat{\mathbb{K}}_1,\dots,\hat{\mathbb{K}}_6\}$:
\begin{equation}\label{ruleTI}
\mathbb{K}_i=\sum_{n=1}^6a_{in}\hat{\mathbb{K}}_n\qquad\text{for}\qquad i=\{1,\dots,6\},
\end{equation}
such that 
\begin{equation}\label{genSymmAdd-TI}
\sum_{n=1}^6\mathbb{K}_n=\sum_{n=1}^6\sum_{i=1}^6a_{in}\mathbb{\hat{K}}_n=\mathbb{S}.
\end{equation} 
From Eq. \eqref{ruleTI} we know that now the problem has 36 unknowns, i.e. the coefficients $a_{in}$ (with $i,n=\{1,\dots,6\}$) and from Eq. \eqref{genSymmAdd-TI} we obtain only 7 equations. The number of equations follows from the symmetries of the TI tensors. Indeed, the TI bases $\hat{\mathbb{K}}_n$ have only the minor symmetries (see Eq. \eqref{symmS} in Section 2). Therefore the 7 equations can be obtained from the components $\{1111\},\{1122\},\{1133\},\{3311\},\{3333\},\{1212\}$ and $\{1313\}$ of Eq. \eqref{genSymmAdd-TI}.
To obtain the remaining equations we cannot impose idempotence and orthogonality since for TI tensors these two properties are incompatible, meaning that a set of TI bases cannot be simultaneously idempotent and orthogonal. The main conclusion here is: Method 1 only tells us that for TI tensors there are infinitely many symmetrically additive sets of bases that are a linear combination of a non-symmetrically additive set; however, Method 1 does not yield any explicit expression. We therefore construct the following alternative method:
\begin{theo}\label{M2-TI}
We first write the symmetrically additive bases as follows: \begin{equation}\label{rule-M2-part2}
\mathbb{K}_n=\mathbb{A}:\hat{\mathbb{K}}_n.
\end{equation}
where $\mathbb{A}$ is a fourth-order tensor whose entries are unknown coefficients to be determined. We then impose the additive symmetry property on the bases $\mathbb{K}_n$ by requiring that:
\begin{equation}\label{rule-M2-part1}
\sum_{n=1}^6 \mathbb{K}_n=\mathbb{A}:\left(\sum_{n=1}^6 \hat{\mathbb{K}}_n\right)=\mathbb{S}.
\end{equation}
Since $\mathbb{S}$ is the fourth-order symmetric identity tensor, it follows from Eq. \eqref{rule-M2-part1} that:
\begin{equation}
\mathbb{A}=\left(\sum_{n=1}^{6}\hat{\mathbb{K}}_n\right)^{-1}.
\end{equation}
Therefore, $\mathbb{A}$ is the inverse of the tensor $\sum_{n=1}^6 \hat{\mathbb{K}}_n$.
\end{theo}
Now, if all the 6 bases $\hat{\mathbb{K}}_n$ have the minor symmetries of the tensor $\mathbb{S}$, the tensor given by the sum of the bases also has these symmetries and therefore it is possible to convert it into a 6x6 symmetric matrix $\textbf{A}^{\text{M}}$, which, according to the Mandel-Voigt notation \cite{helnwein2001some}, has the following components: 
\begin{equation}
\textbf{A}^{\text{M}}=
\left(\begin{array}{cccccc}
A_{1111}&A_{1122}&A_{1133}&0&0&0\\
A_{1122}&A_{2222}&A_{2233}&0&0&0\\
A_{1133}&A_{2233}&A_{3333}&0&0&0\\
0&0&0&A_{1313}&0&0\\
0&0&0&0&A_{2323}&0\\
0&0&0&0&0&A_{1212}\\
\end{array}\right).
\end{equation}
We note that, in order to be invertible, a symmetric matrix needs to be positive definite. We omit the proof here; however, it is possible to show that this is true for all TI tensors $\hat{\mathbb{K}}_n$. Moreover, the matrix $\textbf{A}^{\text{M}}$ is unique. This does not mean that there is only one set of symmetrically additive bases that can be derived from a given set of non-symmetrically additive bases; indeed, as discussed above, there is an infinite number. However, we are not interested in finding a specific set, as long as it satisfies the property of additive symmetry.

\subsection{Example 1: Hill bases}
We now apply Method \ref{M2-TI} to obtain a symmetrically additive set of TI bases from the Hill bases in Eq. \eqref{set-TI-Hill}. To simplify the calculations, we consider the fibres to be aligned with the vector $\m=(0,0,1)^{\Tv}$. We take the fibre vector to be one of the reference axes in three-dimensional (3D) Euclidean space. The fibre vector $\m$ is a unit vector. Moreover, in the small deformation setting the vector $\m$ is independent of the deformation. We first obtain the tensor $\mathbb{A}$ which is given by: 
\begin{equation}\label{A-Hill}
\mathbb{A}=\left(\sum_{n=1}^6\hat{\mathbb{H}}_n\right)^{-1}=\left(\mathbb{S}+\boldsymbol{\Theta}\otimes\textbf{m}\otimes\textbf{m}+\textbf{m}\otimes\textbf{m}\otimes \boldsymbol{\Theta}\right)^{-1}.
\end{equation}
 To calculate the inverse, we convert the tensor on the right of Eq. \eqref{A-Hill} into its Mandel-Voigt form. Then we invert the 6x6 symmetric matrix. The result is:
\begin{equation}
 \textbf{A}^{\text{M}}=
\left(\begin{array}{cccccc}
0&-1&1&0&0&0\\
-1&0&1&0&0&0\\
1&1&-1&0&0&0\\
0&0&0&1&0&0\\
0&0&0&0&1&0\\
0&0&0&0&0&1\\
\end{array}\right).
\end{equation}
We can now revert the matrix $\textbf{A}^{\text{M}}$ into a fourth-order tensor and calculate the symmetrically additive bases by using Eq. \eqref{rule-M2-part2}:
\begin{equation}\label{symm-add-Hill}
\begin{split}
&\mathbb{H}_1=\mathbb{A}:\hat{\mathbb{H}}_1=\dfrac{1}{2}\The\otimes(3 \m\otimes\m-\textbf{I}),\qquad \qquad\mathbb{H}_2=\mathbb{A}:\hat{\mathbb{H}}_2=\The\otimes(\textbf{I}-2 \m\otimes\m),\\
&\mathbb{H}_3=\mathbb{A}:\hat{\mathbb{H}}_3=(\m\otimes\m)\otimes(3 \m\otimes\m-\textbf{I}),\qquad
\mathbb{H}_4=\mathbb{A}:\hat{\mathbb{H}}_4=(\m\otimes\m)\otimes(\textbf{I}-2 \m\otimes\m),\\
&\mathbb{H}_5=\mathbb{A}:\hat{\mathbb{H}}_5=\hat{\mathbb{H}}_5,\quad \text{and}\quad \mathbb{H}_6=\mathbb{A}:\hat{\mathbb{H}}_6=\hat{\mathbb{H}}_6,
\end{split}
\end{equation}
where now $\sum_{n=1}^6\mathbb{H}_n=\mathbb{S}$. The split in Eq. \eqref{symm-add-Hill} is valid for any unit vector $\m$. We only used the specific case $\m=\textbf{e}_3$ to simplify the algebraic calculations.
\subsection{Example 2: Spencer bases}
We now consider the set of bases proposed by Spencer. This set leads to a formulation of the constitutive equation which is frequently used in the literature and makes use of constitutive parameters which are easily measurable experimentally. The tensor $\mathbb{A}$ for the Spencer set is given by:
\begin{equation}\label{A-Spencer}
\mathbb{A}=\left(\sum_{n=1}^6\hat{\mathbb{J}}_n\right)^{-1}=\Big(\mathbb{S}+(\textbf{I}+\textbf{m}\otimes\textbf{m})\otimes(\textbf{I}+\textbf{m}\otimes\textbf{m})\Big)^{-1}
\end{equation}
which can be written in the following Mandel-Voigt form:
\begin{equation}\label{AM-Spencer}
\textbf{A}^{\text{M}}=
\left(\begin{array}{cccccc}
\nicefrac{6}{7}&-\nicefrac{1}{7}&-\nicefrac{2}{7}&0&0&0\\
-\nicefrac{1}{7}&\nicefrac{6}{7}&-\nicefrac{2}{7}&0&0&0\\
-\nicefrac{2}{7}&-\nicefrac{2}{7}&\nicefrac{3}{7}&0&0&0\\
0&0&0&1&0&0\\
0&0&0&0&1&0\\
0&0&0&0&0&1\\
\end{array}\right).
\end{equation} 
The symmetrically-additive set is then given by:
\begin{equation}\label{symm-add-Spencer}
\begin{split}
&\mathbb{J}_1=\mathbb{A}:\hat{\mathbb{J}}_1=\dfrac{1}{7} (3 \textbf{I}- 4 \textbf{m}\otimes\textbf{m})\otimes\textbf{I},\qquad \qquad\mathbb{J}_2=\mathbb{A}:\hat{\mathbb{J}}_2=\dfrac{1}{7} (3 \textbf{I} - 4 \textbf{m}\otimes\textbf{m})\otimes\textbf{m}\otimes\textbf{m},\\
&\mathbb{J}_3=\mathbb{A}:\hat{\mathbb{J}}_3=-\dfrac{1}{7} (2 \textbf{I}-5 \textbf{m}\otimes\textbf{m})\otimes\textbf{I},\hspace{3.2em}
\mathbb{J}_4=\mathbb{A}:\hat{\mathbb{J}}_4=-\dfrac{1}{7} (2 \textbf{I} -5\textbf{m}\otimes\textbf{m})\otimes\textbf{m}\otimes\textbf{m},\\
&\mathbb{J}_5=\mathbb{A}:\hat{\mathbb{J}}_5=\hat{\mathbb{J}}_5-\dfrac{1}{7} ( \textbf{I}+\textbf{m}\otimes\textbf{m}) (\textbf{I} - 
   3 \textbf{m}\otimes\textbf{m},\\
&\mathbb{J}_6=\mathbb{A}:\hat{\mathbb{J}}_6=\hat{\mathbb{J}}_6-\dfrac{4}{7} (\textbf{I} + \textbf{m}\otimes\textbf{m})\otimes(\textbf{m}\otimes\textbf{m}).
\end{split}
\end{equation}
Let us now write down the constitutive equation for a TI linear viscoelastic material. We use the symmetrically additive set in \eqref{symm-add-Hill} to split the relaxation tensor $\mathbb{G}$ and the Spencer set in \eqref{set-TI-1} to write the elastic stress. This choice will allow us to simplify the calculations; however, we could alternatively use the symmetrically additive set in \eqref{symm-add-Spencer} to split $\mathbb{G}$ and to obtain the same constitutive equation. We leave this as exercise for the interested reader. By making use of the contraction products in Table \ref{tab:contr},
\begin{table}
\centering
\begin{tabular}{c|c|c|c|c|c|c}
:&$\hat{\mathbb{J}}_1$
&$\hat{\mathbb{J}}_2$
&$\hat{\mathbb{J}}_3$
&$\hat{\mathbb{J}}_4$
&$\hat{\mathbb{J}}_5$
&$\hat{\mathbb{J}}_6$\\ \hline
$\mathbb{H}_1$
&0
&0
&$\hat{\mathbb{J}}_1-\hat{\mathbb{J}}_3$
&$\hat{\mathbb{J}}_2-\hat{\mathbb{J}}_4$
&$\nicefrac{1}{2}(\hat{\mathbb{J}}_3+\hat{\mathbb{J}}_4-\hat{\mathbb{J}}_1-\hat{\mathbb{J}}_2)$
&$2(\hat{\mathbb{J}}_2-\hat{\mathbb{J}}_4)$\\ \hline
$\mathbb{H}_2$
&$\hat{\mathbb{J}}_1-\hat{\mathbb{J}}_3$
&$\hat{\mathbb{J}}_2-\hat{\mathbb{J}}_4$
&$\hat{\mathbb{J}}_3-\hat{\mathbb{J}}_1$
&$\hat{\mathbb{J}}_4-\hat{\mathbb{J}}_2$
&$\hat{\mathbb{J}}_1-\hat{\mathbb{J}}_3$
&$2(\hat{\mathbb{J}}_4-\hat{\mathbb{J}}_2)$\\ \hline
$\mathbb{H}_3$
&0
&0
&$2\hat{\mathbb{J}}_3$
&$2\hat{\mathbb{J}}_4$
&$-\hat{\mathbb{J}}_3-\hat{\mathbb{J}}_4$
&$4\hat{\mathbb{J}}_4$\\ \hline
$\mathbb{H}_4$
&$\hat{\mathbb{J}}_3$
&$\hat{\mathbb{J}}_4$
&$-\hat{\mathbb{J}}_3$
&$-\hat{\mathbb{J}}_4$
&$\hat{\mathbb{J}}_3$
&$-2\hat{\mathbb{J}}_4$\\ \hline
$\mathbb{H}_5$
&0
&0
&0
&0
&$\hat{\mathbb{J}}_5+\nicefrac{1}{2}(\hat{\mathbb{J}}_2+\hat{\mathbb{J}}_3
+\hat{\mathbb{J}}_4-\hat{\mathbb{J}}_1)$
&0\\ \hline
$\mathbb{H}_6$
&0
&0
&0
&0
&0
&$\hat{\mathbb{J}}_6-2\hat{\mathbb{J}}_4$\\ \hline
\end{tabular}\caption{Contraction table between the symmetrically additive Hill set $\mathbb{H}_n$ and the non-symmetrically additive Spencer set $\hat{\mathbb{J}}_n$.}\label{tab:contr}
\end{table}
we obtain:
\begin{equation}\label{sig-VE-TI-splitM2}
\begin{split}
\sig(t)&=\int_0^t\mathbb{G}(t-\tau):\dfrac{\dd\sig^{\el}(\tau)}{\dd \tau}\dd\tau=\int_0^t\mathbb{G}(t-\tau):\dfrac{\dd\left(\mathbb{C}^{\ti}:\eps(\tau)\right)}{\dd \tau}\dd\tau\\
&=\int_0^t\sum_{n=1}^6G_n(t-\tau)\mathbb{H}_n:\dfrac{\dd\left(\sum_{N=1}^6\hat{C}_N^{\ti}\hat{\mathbb{J}}_N:\eps(\tau)\right)}{\dd\tau}\dd\tau\\
&=\int_0^t\sum_{n=1}^6\sum_{N=1}^6G_n(t-\tau)\hat{C}_N^{\ti}\mathbb{H}_n:\hat{\mathbb{J}}_N:\dfrac{\dd\eps(\tau)}{\dd\tau}\dd\tau\\
&=-\int_0^t\Big(\Delta\Big(D\,G_1(t-\tau)+B\,G_2(t-\tau)\Big)+\mu_tG_5(t-\tau)\Big)\hat{\mathbb{J}}_1:\dfrac{\dd\eps(\tau)}{\dd\tau}\dd\tau\\
&+\int_0^t\Big(\Delta\Big(C\,G_1(t-\tau)+A\,G_2(t-\tau)\Big)+\mu_t\,G_5(t-\tau)\Big)\hat{\mathbb{J}}_2:\dfrac{\dd\eps(\tau)}{\dd\tau}\dd\tau\\
&+\int_0^t\Big(\Delta\Big(B(G_2(t-\tau)-G_4(t-\tau))+D(G_1(t-\tau)-2G_3(t-\tau))\Big)+\mu_t\,G_5(t-\tau)\Big)\hat{\mathbb{J}}_3:\dfrac{\dd\eps(\tau)}{\dd\tau}\dd\tau\\
&-\int_0^t\Big(\Delta\Big(A\big(G_2(t-\tau)-G_4(t-\tau)\big)+C\big(G_1(t-\tau)-2G_3(t-\tau)\Big)\\
&-\mu_t\,G_5(t-\tau)+4\mu_l\,G_6(t-\tau)\Big)\hat{\mathbb{J}}_4:\dfrac{\dd\eps(\tau)}{\dd\tau}\dd\tau\\
&+\int_0^t2\mu_tG_5(t-\tau)\hat{\mathbb{J}}_5:\dfrac{\dd\eps(\tau)}{\dd\tau}\dd\tau+\int_0^t2\mu_lG_6(t-\tau)\hat{\mathbb{J}}_6:\dfrac{\dd\eps(\tau)}{\dd\tau}\dd\tau\\
&=\int_0^t\sum_{n=1}^6R_n(t-\tau)\hat{\mathbb{J}}_n:\dfrac{\dd\,\eps(\tau)}{\dd\tau}\dd\tau,
\end{split}
\end{equation}
where:
\begin{equation}
\begin{split}
&A= \frac{1}{\Delta}(\alpha-\beta-4\mu_L), \quad B= \frac{1}{\Delta}(\alpha-\lambda-2\mu_T),\quad C = \frac{1}{\Delta}(\beta+4\mu_L-\mu_T), \quad D= \frac{1}{\Delta}(\mu_T-\alpha),\\
&\Delta = (\alpha-\lambda-2\mu_T)(\beta+4\mu_L-\mu_T)-(\alpha-\beta-4\mu_L)(\mu_T-\alpha),
\end{split}
\end{equation}
and
\begin{equation}\label{relax-Rn}
R_1(t)=\lambda(t),\qquad R_2(t)=R_3(t)=\alpha(t),\qquad R_4(t)=\beta(t),\qquad R_5(t)=2\mu_t(t),\qquad
R_6(t)=2\mu_l(t).\end{equation}
By comparing the last two equations in \eqref{sig-VE-TI-splitM2} we find the links between the reduced relaxation functions $G_n(t)$ and the time-dependent moduli $\lambda(t),\alpha(t),\beta(t),\mu_t(t)$ and $\mu_l(t)$, which are given by:
\begin{equation}\label{conv-Hill-symm-Spencer}
\begin{split}
&G_1(t)= A\Big(\lambda(t)+\mu_t(t)\Big)+B \Big(\alpha(t)-\mu_t(t)\Big),\qquad
G_2(t)=  -C\Big(\lambda(t)+\mu_t(t)\Big)-D\Big(\alpha(t)-\mu_t(t)\Big),\\
&G_3(t)=\dfrac{A}{2}\Big(\lambda(t)+\alpha(t)\Big)+\dfrac{B}{2}\Big(\alpha(t)+\beta(t)-2\mu_t(t)+4\mu_l(t)\Big),\\
&G_4(t)=-C\Big(\lambda(t)+\alpha(t)\Big)-D\Big(\alpha(t)+\beta(t)+2\mu_t(t)-4 \mu_l(t)\Big),\quad
G_5(t)= \frac{\mu_t(t)}{\mu_t},\quad G_6(t)= \frac{\mu_l(t)}{\mu_l}.
\end{split}
\end{equation}
We note that to go from the third to the fourth equation in \eqref{sig-VE-TI-splitM2} we have used the fact that the bases are time-independent (they only depend on the fibres vector $\m$, which in the small deformation regime is a unit vector and does  not depend on the time nor the deformation). \par
The links in \eqref{conv-Hill-symm-Spencer} allow us to find a connection between the functions $G_n(t)$ (associated with the symmetrically additive set of Hill bases $\mathbb{H}$) and the moduli $\lambda(t),\alpha(t),\beta(t),\mu_t(t),\mu_l(t)$, which are associated with 5 physical modes of deformation and therefore can be experimentally measured by performing a hydrostatic compression, a simple shear in the plane of isotropy, two simple shears in the planes perpendicular to the preferred direction, both parallel and perpendicular to the preferred direction, and a uni-axial tension.
\par To summarise the results of this section, we have now constructed a set of symmetrically additive bases to split the relaxation tensor $\mathbb{G}(t)$ for TI viscoelastic materials. In the next section, we will apply the symmetrically additive splits for isotropic and TI tensors to the context of MQLV. We will show that, in the large deformation regime, the bases depend on the deformation and therefore the relaxation tensor $\mathbb{G}$ is naturally deformation-dependent. We will illustrate this property with simple examples, by considering different deformation modes that are typically used in mechanical testing.

\section{Modified quasi-linear viscoelasticity}\label{sec:mQLV}
In the large deformations setting, the deformation of a continuum body is modelled as a linear function $\boldsymbol{\chi}$ (in 3D Euclidean space) that transforms points from an initial undeformed state $\mathcal{B}_0$ to a final deformed state $\mathcal{B}(t)$. We use the notation $\textbf{X}$ and $\textbf{x}(t)$ to denote position vectors in the undeformed and deformed configurations respectively, so that $\textbf{x}(t)=\boldsymbol{\chi}(t,\textbf{X})$. The gradient of the deformation is the tensor $\textbf{F}(t)=\dfrac{\partial \textbf{x}(t)}{\partial \textbf{X}}$ that has components $F_{ab}(t)=\dfrac{\partial x_a(t)}{\partial X_b}$. We assume that the deformation starts at $t=0$ so that $\textbf{F}(t)=\textbf{0},\forall t\leq0$. QLV theory essentially extends the linear visco-elastic theory to the large deformation regime and allows one to account for the non-linear elastic stress-strain response of a material by writing the constitutive equation with respect to the elastic stress instead of the strain tensor. In order to satisfy objectivity, the general form of the constitutive equation is written with respect to the second Piola-Kirchhoff stress $\boldsymbol{\Pi}(t)$. To obtain the modified QLV model (MQLV) in Eq. \eqref{MQLV-Fung}, we start from Eq. \eqref{QLV-Fung} and we integrate by parts:
\begin{equation}\label{Fung-to-mQLV}
\begin{split}
\boldsymbol{\Pi}(t)&=\sum_{n=1}^NG_n(0)\mathbb{K}_n(t):\Big(J(t)\f^{-1}(t)\textbf{T}^{\el}(t)\f^{-\tra}(t)\Big)\\
&+\int_0^t\sum_{n=1}^N G_n'(t-\tau)\mathbb{K}_n(\tau):\Big(J(\tau)\f^{-1}(\tau)\textbf{T}^{\el}(\tau)\f^{-\tra}(\tau)\Big)\dt\\
&-\int_0^t\sum_{n=1}^N G_n(t-\tau)\mathbb{K}_n'(\tau):\Big(J(\tau)\f^{-1}(\tau)\textbf{T}^{\el}(\tau)\f^{-\tra}(\tau)\Big)\dt.
\end{split}
\end{equation}
The MQLV theory applies the bases $\mathbb{K}_n(t)$ directly to the elastic Cauchy stress $\textbf{T}^{\el}$, as follows:
\begin{equation}\label{mQLV}
\begin{split}
\boldsymbol{\Pi}(t)&=\sum_{n=1}^NG_n(0)\boldsymbol{\Pi}_n^{\el}(t)
+\int_0^t\sum_{n=1}^N G_n'(t-\tau)\boldsymbol{\Pi}_n^{\el}(\tau)\dt
-\int_0^t\sum_{n=1}^N G_n(t-\tau)\boldsymbol{\Omega}_n^{\el}(t)\dt
\end{split}
\end{equation}
where we have defined the following terms:
\begin{equation}
\boldsymbol{\Pi}^{\el}_n=J\f^{-1}\textbf{T}_n^{\el}\f^{-\tra}=J\f^{-1}(\mathbb{K}_n:\textbf{T}^{\el})\f^{-\tra}\qquad\text{and}\qquad\boldsymbol{\Omega}^{\el}_n=J\f^{-1}\boldsymbol{\omega}_n^{\el}\f^{-\tra}=J\f^{-1}(\mathbb{K}'_n:\textbf{T}^{\el})\f^{-\tra}.
\end{equation}
The term outside of the integrals gives the elastic second Piola-Kirchhoff stress, provided that the set of bases $\mathbb{K}_n$ is symmetrically additive and the components of $\mathbb{G}$ satisfy the condition $G_n(0)=1,\,\forall n\in\{1,\dots,N\}$. We remark that the symmetrically additive property is crucial in order for the constitutive equation to be consistent with the elastic limit. Moreover, we note that the two integrals in Eq. \eqref{mQLV} arise from the fact that, in the general formulation, the bases $\mathbb{K}_n(t)$ depend on the time $t$. \par
Finally, we recall that the MQLV model is not derived mathematically from the original QLV form proposed by Fung. Indeed, although the MQLV model formally resembles the QLV model, they are two different constitutive equations. The main difference is in the bases that split the tensorial relaxation function: in the QLV model the bases must be written with respect to the undeformed fibre vector $\textbf{M}$, whereas in the MQLV model, the bases must be written with respect to the deformed fibre vector ($\m$ or else its normalised counterpart $\hat{\m}$) in order for the constitutive equation to be objective. \par
In the next sections, we will show that in the isotropic setting, the bases are time-independent, see Eqs. \eqref{set1-iso}, therefore the second integral in Eq. \eqref{mQLV} is identically zero. However, in the transversely isotropic setting, the bases do depend on time and the deformation in general. We will illustrate this property by considering specific deformations.

\subsection{Isotropy}
The isotropic case was extensively studied in \cite{de2014nonlinear}. For completeness we recall here the compressible and incompressible forms. By using the Piola transformation \cite{ogden1997non}, the Cauchy stress for a compressible material can be written as:
\begin{equation}\label{Cauchy}
\begin{split}
\textbf{T}(t)
=&\textbf{T}^{\el}_1(t)+
J^{-1}(t)\f(t)\left(\int_0^t \dfrac{\kappa'(t-\tau)}{\kappa}J(\tau)\f^{-1}(\tau)\textbf{T}_1^{\el}(\tau)\f^{-\tra}(\tau)\dt\right)\f^{\tra}(t)\\
+&\textbf{T}_2^{\el}(t)+J^{-1}(t)\f(t)\left(\int_0^t \dfrac{\mu'(t-\tau)}{\mu}J(\tau)\f^{-1}(\tau)\textbf{T}_2^{\el}(\tau)\f^{-\tra}(\tau)\dt\right)\f^{\tra}(t),
\end{split}
\end{equation}
where:
\begin{equation}\label{T1-T2-iso}
\textbf{T}_1^{\el}=\dfrac{1}{3}\tr[\textbf{T}^{\el}]\textbf{I}\qquad \text{and}\qquad \textbf{T}_2^{\el}=\textbf{T}^{\el}-\dfrac{1}{3}\tr[\textbf{T}^{\el}]\textbf{I}=\dev[\textbf{T}^{\el}],
\end{equation}
and we have replaced $\mathbb{K}_n$ with the symmetrically additive bases in Eq. \eqref{set-TI-1} and the components $G_1$ and $G_2$ with the connections in Eq. \eqref{relax-func-iso}.
\par
\begin{figure}[b!]
\centering
\includegraphics[scale=0.5]{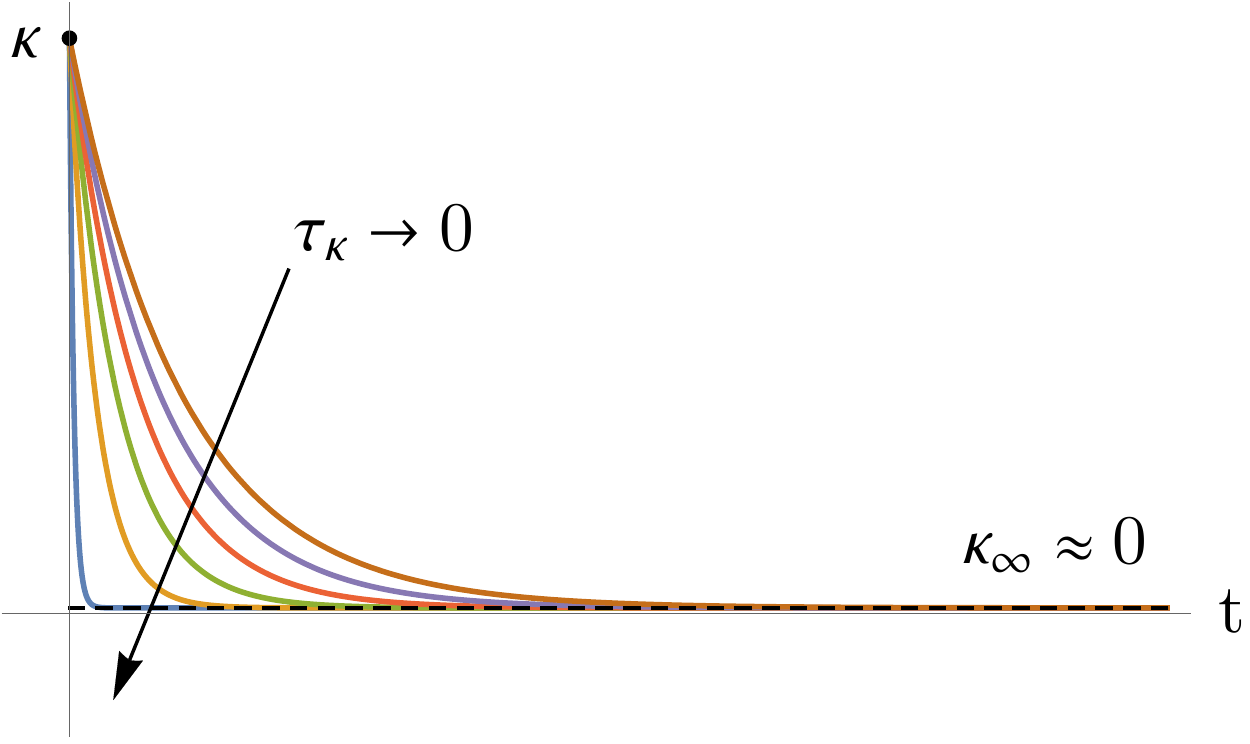}\caption{Behaviour of the relaxation function $\kappa(t)$ as the characteristic time $\tau_{\kappa}\rightarrow0$. The long-term equilibrium bulk modulus is set to $\kappa_{\infty}=0.01$ and the curves are plotted for $\tau_{\kappa}=\{0.52,0.42,0.32,0.22,0.12,0.02\}$.}\label{fig:kappa}
\end{figure}

 We assume a Prony series form for the relaxation functions $\kappa(t)$ and $\mu(t)$ such that:
\begin{equation}
\kappa(t)=\kappa_{\infty}+(\kappa-\kappa_{\infty})e^{-\nicefrac{t}{\tau_{\kappa}}}\qquad\text{and}\qquad\mu(t)=\mu_{\infty}+(\mu-\mu_{\infty})e^{-\nicefrac{t}{\tau_{\mu}}}.
\end{equation}
In the incompressible limit, we assume that the elastic bulk modulus is much greater than the elastic shear modulus so that $\kappa\rightarrow\infty$. Moreover, we take the characteristic time $\tau_{\kappa}$ and the long-term bulk modulus  $\kappa_{\infty}$ to be close to zero in the incompressible limit. These assumptions correspond to the incompressible contributions being instantaneous (elastic) only. In Figure \ref{fig:kappa} we 
show the behaviour of the relaxation function associated to the bulk modulus in the incompressible limit. Finally, for isochoric deformations we have $J(t)=1,\,\forall t$. Under this assumptions we can rewrite Eq. \eqref{Cauchy} as follows:
\begin{equation}\label{Cauchy-incompr}
\textbf{T}(t)
=-p(t)\textbf{I}+\dev[\textbf{T}^{\el}(t)]+\f(t)\left(\int_0^t \dfrac{\mu'(t-\tau)}{\mu}\f^{-1}(\tau)\dev[\textbf{T}^{\el}(\tau)]\f^{-\tra}(\tau)\dt\right)\f^{\tra}(t),
\end{equation} 
where the Lagrange multiplier $p(t)$ is given by:
\begin{equation}\label{p-limit}
\begin{split}
-p(t)\textbf{I}=\lim_{\substack{J\to 1 \\ \kappa\to \infty \\ \kappa_{\infty},\tau_{\kappa}\to 0}}&J(t)^{-1}\dfrac{1}{3}\tr[\textbf{T}^{\el}(t)]\textbf{I}\\
&+
J(t)^{-1}\f(t)\left(\dfrac{1}{3}\int_0^t \dfrac{\kappa'(t-\tau)}{\kappa}J^{-1}(\tau)\f^{-1}(\tau)\tr[\textbf{T}^{\el}(\tau)]\f^{-\tra}(\tau)\dt\right)\f^{\tra}(t).
\end{split}
\end{equation}

\subsection{Transverse isotropy}
We call $\textbf{M}$ the unit vector that identifies the preferred direction of the TI material in the initial undeformed configuration and $\m(t)=\textbf{F}(t)\textbf{M}$ the equivalent deformed fibre vector. For a compressible TI material, the constitutive equation for the elastic Cauchy stress is given by:
\begin{equation}\label{const-el-cauchy}
\begin{split}
\textbf{T}^{\el}(t)=2J(t)^{-1}\Big((I_3(t)W_3(t)\textbf{I}&+W_1(t)\textbf{B}(t)- W_2(t)\textbf{B}(t)^{-1}+W_4(t)\m(t)\otimes\m(t)\\
&+W_5(t)\Big(\textbf{B}(t)\m(t)\otimes\m(t)+\m(t)\otimes\textbf{B}(t)\m(t)\Big)\Big),
\end{split}
\end{equation}
where $J(t)=\text{det}\textbf{F}(t)$, $W_i(t)=\nicefrac{\partial W(t)}{\partial I_i(t)}$ ($i=\{1,\dots,5\}~$) and $I_i$ are the invariants of $\textbf{B}(t)$, see \cite{ogden1997non} for the full details. For the strain energy function $W(t)$, we choose the following form:
\begin{equation}
W(t)=W_{\text{iso}}(t)+\dfrac{\mu_t-\mu_l}{2} (2 I_4(t)-I_5(t)-1) +\dfrac{E_l+\mu_t-4 \mu_l}{16} (I_4(t)-1)(I_5(t)-1),
\end{equation}
where
\begin{equation}\label{Wiso}
W_{\text{iso}}(t)=\dfrac{\mu_t}{2}  \big(\alpha_{\text{MR}}  (I_1(t)-3)+(1-\alpha_{\text{MR}}) (I_2(t)-3)\big),\qquad \text{with}\quad \alpha_{\text{MR}}\in[0,1],
\end{equation}
which is consistent with the linear theory in the small strain limit \cite{murphy2013transversely}. \par
Now, to write the MQLV model, we choose the symmetrically additive Hill bases; therefore, the constitutive equation \eqref{mQLV} for compressible TI materials can be written as follows:
\begin{equation}\label{mQLV-TI}
\begin{split}
\textbf{T}(t)&=J^{-1}(t)\f(t)\Big(\sum^{6}_{n=1}R_n(0)\boldsymbol{P}_{n}^{\el}(t)\Big)\f^{\tra}(t)\\
&+J^{-1}\f(t)\Big(\int^{t}_{0}\sum^{6}_{n=1}R_n'(t-\tau)\boldsymbol{P}^{\el}_n(\tau)\text{d}\tau
-\int^{t}_{0}\sum^{6}_{n=1}R_n(t-\tau)\boldsymbol{Q}^{\el}_n(\tau)\text{d}\tau\Big)\f^{\tra}(t),
\end{split}
\end{equation}
where we have used the connection in Eq. \eqref{conv-Hill-symm-Spencer} to replace the components $G_n$, with the relaxation functions $R_n$, $n=\{1,\dots,6\}$. The terms $\boldsymbol{P}_n^{\el}$ and $\boldsymbol{Q}_n^{\el}$ are given by:
\begin{equation}\label{bigP}
\begin{split}
\boldsymbol{P}^{\el}_1&= A\Big(\PPi^{\el}_1+\dfrac{\PPi^{\el}_3}{2}\Big)-C\Big(\PPi^{\el}_2+\PPi^{\el}_4\Big)=J\left(A\,\tilde{T}^{\el}-C\,\bar{T}^{\el}\right)\C^{-1},\\
\boldsymbol{P}^{\el}_2&=B\Big(\PPi^{\el}_1+\dfrac{\PPi^{\el}_3}{2}\Big)- D\Big(\PPi^{\el}_2+\PPi^{\el}_4\Big)=J\left(B\,\tilde{T}^{\el}-D\,\bar{T}^{\el}\right)\C^{-1},\\
\boldsymbol{P}^{\el}_3&=\nicefrac{A}{2}\,\PPi^{\el}_3-C\,\PPi^{\el}_4=J\left(A\,\tilde{T}^{\el}-C\,\bar{T}^{\el}\right)\f^{-1}\m\otimes\f^{-1}\m,\\
\boldsymbol{P}^{\el}_4&=\nicefrac{B}{2}\,\PPi^{\el}_3- D\,\PPi^{\el}_4=J\left(B\,\tilde{T}^{\el}-D\,\bar{T}^{\el}\right)\f^{-1}\m\otimes\f^{-1}\m,\\
\boldsymbol{P}^{\el}_5&=A\,\PPi^{\el}_1-B\left(\PPi^{\el}_1+\PPi^{\el}_3\right)-C\,\PPi^{\el}_2+D\left(2\PPi^{\el}_4+\PPi^{\el}_2\right)+\dfrac{\PPi^{\el}_5}{\mu_t}\\
&=J\big(A\tilde{T}^{\el}-C\bar{T}^{\el}\big)\left(\textbf{C}^{-1}-\f^{-1}\m\otimes\f^{-1}\m\right)-J\big(B\tilde{T}^{\el}-D\bar{T}^{\el}\big)\left(\C^{-1}+\f^{-1}\m\otimes\f^{-1}\m\right)+\dfrac{\PPi^{\el}_5}{\mu_t},\\
\boldsymbol{P}^{\el}_6&=2(B\,\PPi^{\el}_3 - 2D\, \PPi^{\el}_4) + \dfrac{\PPi^{\el}_6}{\mu_l}
=4J\left(B\,\tilde{T}^{\el}-D\,\bar{T}^{\el}\right)\f^{-1}\m\otimes\f^{-1}\m+ \dfrac{\PPi^{\el}_6}{\mu_l},
\end{split}
\end{equation}
and
\begin{equation}\label{bigQ}
\begin{split}
\boldsymbol{Q}^{\el}_1&= A\Big(\Om_1+\dfrac{\Om_3}{2}\Big)-C\Big(\Om_2+\Om_4\Big)=J\left(\nicefrac{3}{2} A+2C\right)\tpar\C^{-1},\\
\boldsymbol{Q}^{\el}_2&=B\Big(\Om_1+\dfrac{\Om_3}{2}\Big)- D\Big(\Om_2+\Om_4\Big)=J\left(\nicefrac{3}{2} B+2D\right)\tpar\C^{-1},\\
\boldsymbol{Q}^{\el}_3&=\nicefrac{A}{2}\,\Om_3-C\,\Om_4=J\left(A\,\tilde{T}^{\el}-C\,\bar{T}^{\el}\right)\left(\f^{-1}\m'\otimes\f^{-1}\m+\f^{-1}\m\otimes\f^{-1}\m'\right)
\\&+J\left(\nicefrac{3}{2} A+2C\right)\tpar\,\f^{-1}\m\otimes\f^{-1}\m,\\
\boldsymbol{Q}^{\el}_4&=\nicefrac{B}{2}\,\Om_3-D\,\Om_4=J\left(B\,\tilde{T}^{\el}-D\,\bar{T}^{\el}\right)\left(\f^{-1}\m'\otimes\f^{-1}\m+\f^{-1}\m\otimes\f^{-1}\m'\right)
\\&+J\left(\nicefrac{3}{2} B+2D\right)\tpar\,\f^{-1}\m\otimes\f^{-1}\m,\\
\boldsymbol{Q}^{\el}_5&=A\,\Om_1-B\left(\Om_1+\Om_3\right)-C\,\Om_2+D\left(2\Om_4+\Om_2\right)+\dfrac{\Om_5}{\mu_t}\\
&=J\big(\nicefrac{3}{2}A+2C\big)\tpar\left(\C^{-1}-\f^{-1}\m\otimes\f^{-1}\m\right)-J\big(\nicefrac{3}{2}B+2D\big)\tpar\left(\C^{-1}+\f^{-1}\m\otimes\f^{-1}\m\right)
\\&-J\big((A+B)\tilde{T}^{\el}+A\bar{T}^{\el}\big)\left(\f^{-1}\m'\otimes\f^{-1}\m+\f^{-1}\m\otimes\f^{-1}\m'\right)+\dfrac{\Om_5}{\mu_t},\\
\boldsymbol{Q}^{\el}_6&=2(B\,\Om_3 - 2D\, \Om_4) + \dfrac{\Om_6}{2 \mu_l}
=4J\left(B\,\tilde{T}^{\el}-D\,\bar{T}^{\el}\right)\left(\f^{-1}\m'\otimes\f^{-1}\m+\f^{-1}\m\otimes\f^{-1}\m'\right)\\&
+4J(\nicefrac{3}{2}B+2D)\tpar\,\f^{-1}\m\otimes\f^{-1}\m+ \dfrac{\Om_6}{\mu_l},\\
\end{split}
\end{equation}
respectively. We have defined the following terms:
\begin{equation}\label{terms-P-Q}
T_{\|}^{\el}=\m\cdot\textbf{T}^{\el}\m,
\qquad 
\tpar=\m'\cdot\textbf{T}^{\el}\m+\m\cdot\textbf{T}^{\el}\m',
\qquad
\tilde{T}^{\el}=\nicefrac{1}{2}(3T_{\|}^{\el}-\tr[\textbf{T}^{\el}]),
\qquad 
\bar{T}^{\el}=\tr[\textbf{T}^{\el}]-2T_{\|}^{\el}.
\end{equation}
In the incompressible limit, we assume that the relaxation function $\lambda(t)$ behaves as the function $\kappa(t)$ in Figure \ref{fig:kappa}, with $\lambda\to\infty$ and $\tau_{\lambda},\lambda_{\infty}\to0$. Therefore:
\begin{equation}
\lim_{\lambda\to\infty}A=\lim_{\lambda\to\infty}C=\lim_{\lambda\to\infty}D=0 \qquad\text{and}\qquad\lim_{\lambda\to\infty}B=\dfrac{1}{\beta+4\mu_l-\mu_t}=\dfrac{1}{E_l},
\end{equation}
where $E_l$ is the elastic longitudinal Young modulus.
Moreover, in the incompressible limit $\alpha\to\mu_t$; therefore, we take $\alpha(t)\to\mu_t(t),\,\forall t$. Under these assumptions, the constitutive equation \eqref{mQLV-TI} can be rewritten as follows:
\begin{equation}\label{mQLV-TI-incompr}
\begin{split}
\textbf{T}(t)&=-p(t)\textbf{I}+\tilde{T}^{\el}(t)\m(t)\otimes\m(t)+\textbf{T}_5^{\el}(t)+\textbf{T}_6^{\el}(t)\\
&+\int_0^t\dfrac{E_l'(t-\tau)}{E_l}\PPi_L^{\el}(\tau)\dd\tau+\int_0^t\dfrac{E_l(t-\tau)}{E_l}\Om_L(\tau)\dd\tau\\
&+\int_0^t\dfrac{\mu_t'(t-\tau)}{\mu_t}\PPi_5^{\el}(\tau)\dd\tau+\int_0^t\dfrac{\mu_t(t-\tau)}{\mu_t}\Om_T^{\el}(\tau)\dd\tau\\
&+\int_0^t\dfrac{\mu_l'(t-\tau)}{\mu_l}\PPi_6^{\el}(\tau)\dd\tau+\int_0^t\dfrac{\mu_l(t-\tau)}{\mu_l}\Om_A^{\el}(\tau)\dd\tau,
\end{split}
\end{equation}
where:
\begin{equation}\label{Pi-Om-TI}
\begin{split}
\PPi_L^{\el}&=\tilde{T}^{\el}\f^{-1}\m\otimes\m\,\f^{-\tra},
\hspace{6.5em}
\Om_L =\f^{-1}\left(\tilde{T}^{\el}(\m'\otimes\m+\m\otimes\m')+\dfrac{3}{2}\,\tilde{T}^{\el}_{\|}\m\otimes\m\right)\f^{-\tra},\\
\PPi_T^{\el}&=\f^{-1}\textbf{T}_5^{\el}\f^{-\tra},
\qquad
\Om_T=\f^{-1}(\mathbb{K}_5':\textbf{T}^{\el})\f^{-\tra},
\qquad
\PPi_A^{\el}=\f^{-1}\textbf{T}_6^{\el}\f^{-\tra},
\qquad
\Om_A=\f^{-1}(\mathbb{K}_6':\textbf{T}^{\el})\f^{-\tra},
\end{split}
\end{equation}
and $E_l(t)=\beta(t)-\mu_t(t)+4\mu_l(t)$. The Lagrange multiplier $p(t)$ is given by:
\begin{equation}
\begin{split}
-p(t)\textbf{I}=\lim_{\substack{J\to1,\lambda\to\infty,\\ \lambda_{\infty},\tau_{\lambda}\to 0}}&
%
\lambda \left(A\,\tilde{T}^{\el}(t)-C\,\bar{T}^{\el}(t)\right)\textbf{I}\\
+&J^{-1}(t)\f^{-1}(t)\left(\int_0^t\lambda'(t-\tau)J(\tau)\left(A\tilde{T}^{\el}(\tau)-C\bar{T}^{\el}(\tau)\right)\C^{-1}(\tau)\dd\tau\right)\f^{-\tra}(t)\\
+&J^{-1}(t)\f^{-1}(t)\left(\int_0^t\lambda(t-\tau)J(\tau)\left(\nicefrac{3}{2} A+2C\right)\tpar(\tau)\C^{-1}(\tau)\dd\tau\right)\f^{-\tra}(t).
\end{split}
\end{equation}
In the next section, we illustrate the key features of the MQLV model for TI materials. 

\section{Strain-dependent relaxation}
A key feature of the MQLV model for TI materials is that the constitutive equation is able to capture strain-dependent relaxation. This non-linear property is a direct consequence of the bases being dependent on the deformed fibre vector $\m(t)$. The vector $\m(t)$ depends on the deformation through the deformation gradient $\f(t)$. This naturally introduces a dependence on the deformation in the relaxation tensor $\mathbb{G}(t)$. Therefore, the model naturally captures the non-linear phenomenon of strain-dependent relaxation, which  is commonly observed in soft materials such as soft tissues and gels, whereby the relaxation curve is affected by the level of strain reached during a step-strain test \cite{shearer2020recruitment,nasseri2002viscoelastic,safshekan2017viscoelastic,chatelin2010fifty}. \par
\begin{table}[b!]
\centering
\resizebox{\columnwidth}{!}{
\begin{tabular}{|p{19mm}|c|c|c|}
\hline
& Deformation gradient & Deformed fibre vector & Deformed unit fibre vector\\ \hline
Uni-axial & 
$\textbf{F}(t)=\text{diag}\left(\dfrac{1}{\sqrt{\Lambda(t)}},\dfrac{1}{\sqrt{\Lambda(t)}},\Lambda(t)\right)$
& $\m(t)=\left(\begin{array}{c} 
0\\
0\\
\Lambda(t)
\end{array} \right) $ & $\hat{\m}=\left(\begin{array}{c} 
0\\
0\\
1
\end{array} \right) $\\\hline
In-plane shear & 
$\textbf{F}(t)=\left(\begin{array}{ccc} 
1& \kappa_2(t) &0\\
0 & 1 & 0\\
0 & 0 &1
\end{array} \right) $
 & $\m=\left(\begin{array}{c} 
0\\
0\\
1
\end{array} \right) $ 
& $\hat{\m}=\left(\begin{array}{c} 
0\\
0\\
1
\end{array} \right) $\\\hline
Longitudinal \newline shear & 
$\textbf{F}(t)=\left(\begin{array}{ccc} 
1&0 &0\\
0 & 1 & 0\\
\kappa_3(t) & 0 &1
\end{array} \right) $
 & $\m=\left(\begin{array}{c} 
0\\
0\\
1
\end{array} \right) $ 
& $\hat{\m}=\left(\begin{array}{c} 
0\\
0\\
1
\end{array} \right) $\\\hline
Perpendicular shear & 
$\textbf{F}(t)=\left(\begin{array}{ccc} 
1&0 &\kappa_3(t)\\
0 & 1 & 0\\
0 & 0 &1
\end{array} \right) $
 & $\m(t)=\left(\begin{array}{c} 
\kappa_3(t)\\
0\\
1
\end{array} \right) $ 
& $\hat{\m}(t)=\left(\begin{array}{c} 
\dfrac{\kappa_3(t)}{\sqrt{1+\kappa_3(t)^2}}\\
0\\
\dfrac{1}{\sqrt{1+\kappa_3(t)^2}}
\end{array} \right) $\\\hline
\end{tabular}
}
\caption{Deformation gradient $\textbf{F}(t)$, deformed fibre vector $\m(t)=\textbf{F}(t)\textbf{M}$ and normalised fibre vector $\hat{\m}(t)$ from Eq. \eqref{hat-m} for uni-axial elongation along the fibres direction, in-plane shear, longitudinal and perpendicular shear. All the four deformations are illustrated in Figure \ref{fig:fibres}. The undeformed fibre vector is taken to be $\textbf{M}=\textbf{e}_3=(0,0,1)^{\tra}$.}\label{tab:deform-vectors}
\end{table}
\begin{figure}[h!]
\centering
\includegraphics[scale=0.34]{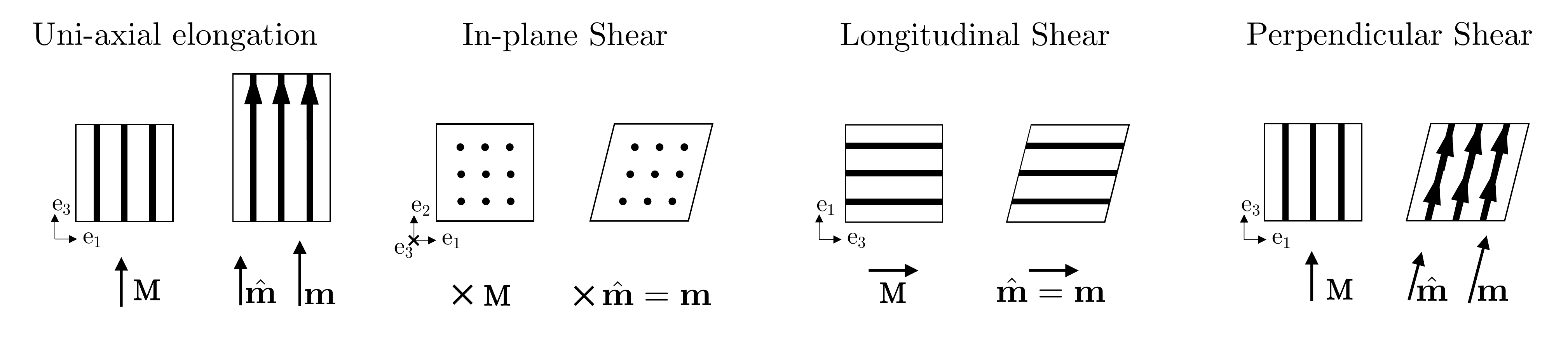}\caption{Deformation modes for a fibre-reinforced TI material: in uni-axial elongation, the fibres stretch ($\hat{\m}=\textbf{M}$ and $\m(t)=(0,0,\Lambda(t))^{\text{T}}$); under in-plane and longitudinal shear, the fibres do not deform ($\hat{\m}=\m=\textbf{M}$); in perpendicular shear, the fibres stretch and rotate ($\hat{\m}(t)\neq\m(t)\neq\textbf{M}$).}\label{fig:fibres}
\centering
\end{figure}
We note that the bases used in the constitutive equation \eqref{mQLV-TI-incompr} are written with respect to the vector $\m(t)$. However, an alternative formulation (which was proposed in \cite{balbi2018modified}) allows us to write the bases with respect to the unit deformed vector:
\begin{equation}\label{hat-m}
\hat{\m}(t)=\dfrac{\m(t)}{\|\m(t)\|}=\dfrac{\f(t)\textbf{M}}{\sqrt{\f(t)\textbf{M}\cdot\f(t)\textbf{M}}}.
\end{equation}
In Table \ref{tab:deform-vectors}, we consider the case of a fibre-reinforced TI material and we compare the vectors $\hat{\m}$ and $\m$ for the most common deformation modes used in mechanical testing. These are: uni-axial extension, in-plane shear (i.e. shear in the isotropic plane), longitudinal and perpendicular shear (i.e. shear along the fibre- and perpendicular to the fibre- direction, respectively). The fibres in the undeformed material are assumed to be aligned along the vector $\textbf{M}=(0,0,1)^{\text{T}}$. The two versions of the MQLV theory are equivalent for deformation modes where $\hat{\m}=\m$, i.e. in-plane and longitudinal shear. For these deformation modes, the fibres do not deform; therefore, $\hat{\m}=\m=\textbf{M}$, as shown in Figure \ref{fig:fibres}. However, they differ under uni-axial extension and perpendicular shear, which we consider below.
\\

\begin{figure}[t!]
\centering
\includegraphics[scale=0.36]{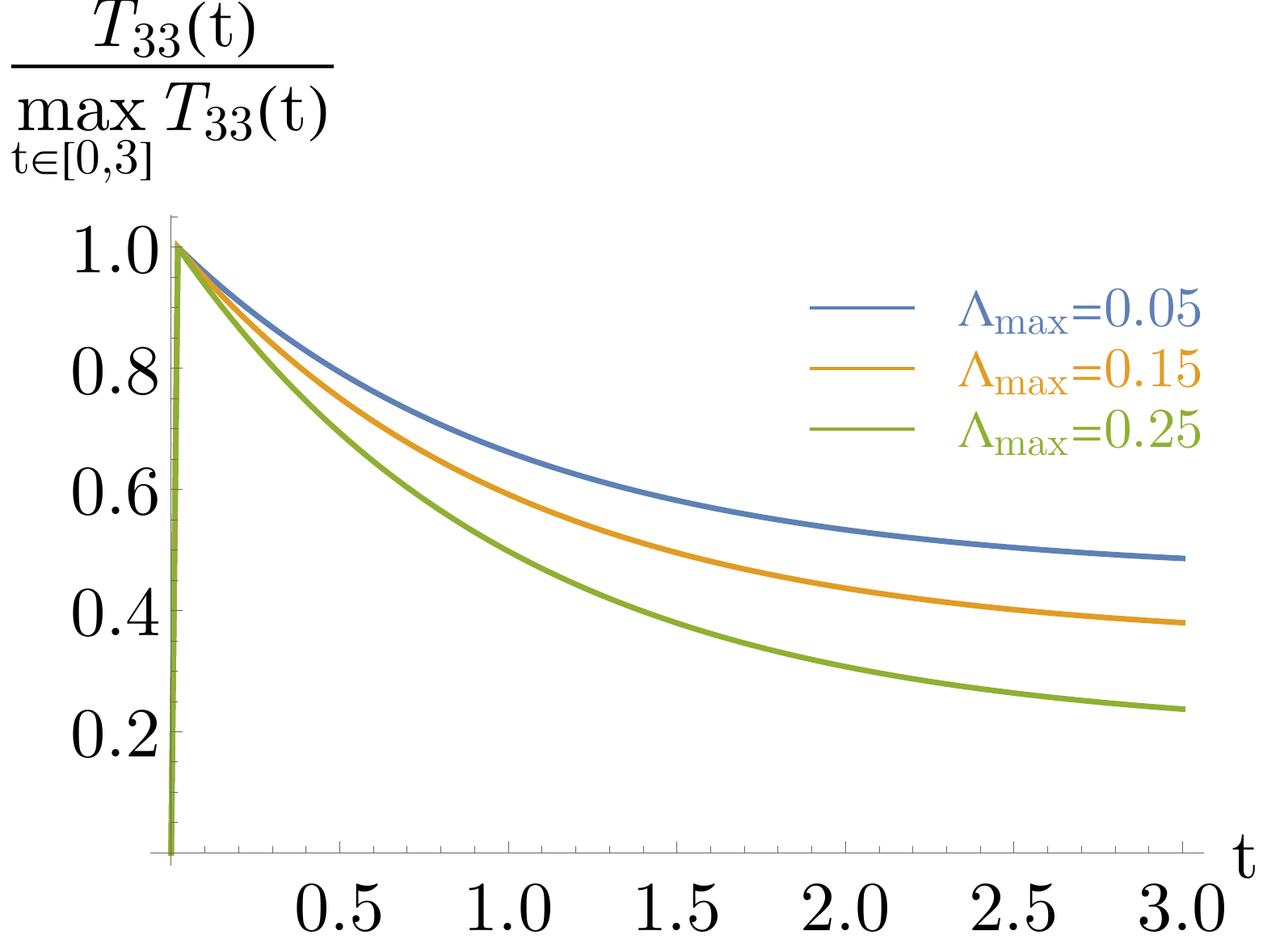}\hfill
\includegraphics[scale=0.36]{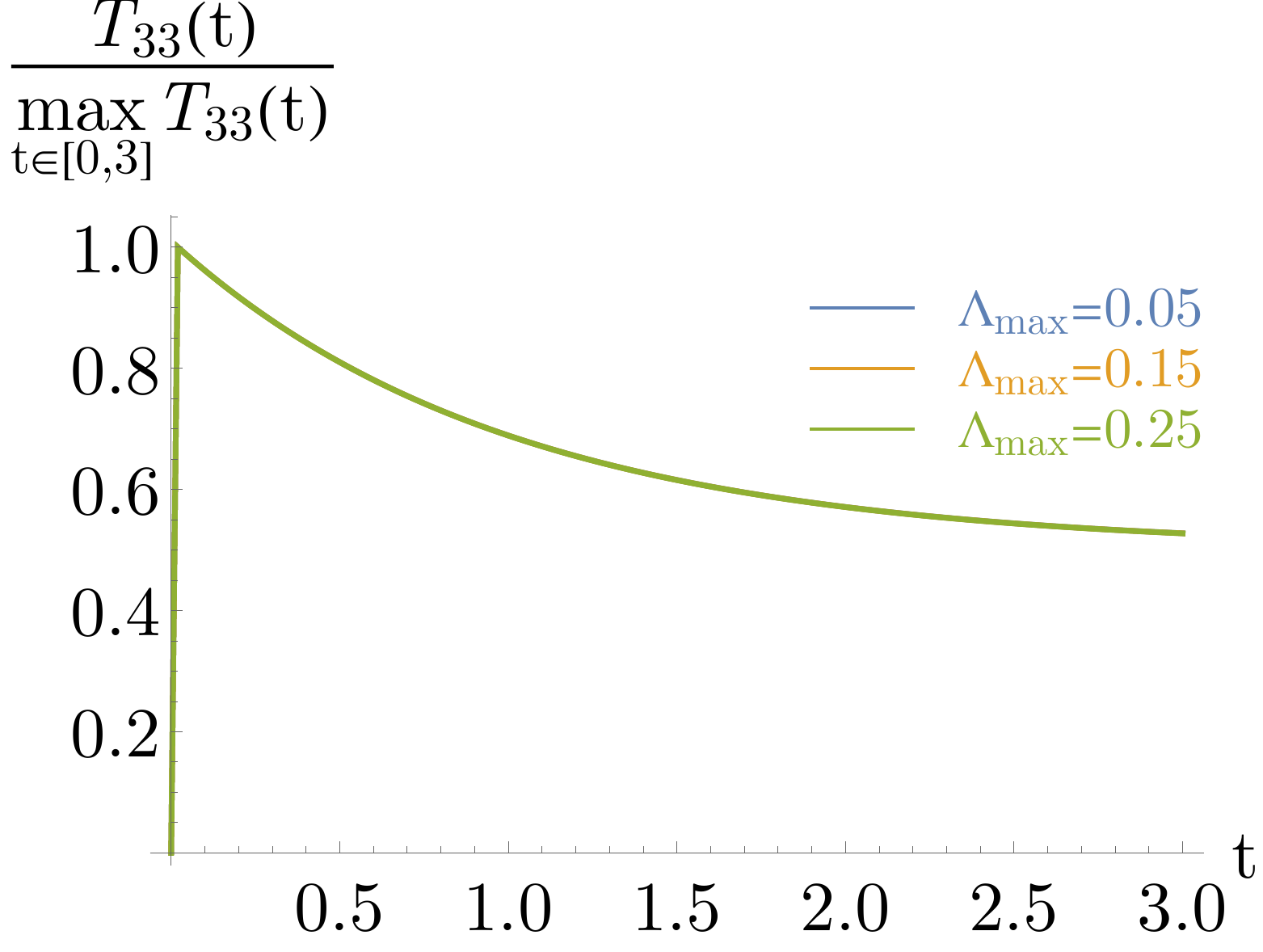} \caption{\textbf{Uni-axial elongation}. Normalised stress relaxation curves for the stress $T_{33}(t)$ using the MQLV formulation written with respect to $\m$ (left) and $\hat{\m}$ (right) at different level of strain $\Lambda_{\text{max}}=\{0.05,0.15,0.25\}$. The curves are plotted by using Eqs. \eqref{uniaxial-m(t)} and \eqref{T33hatm}, respectively. The stretch $\Lambda(t)$ is in the form of a ramp function. The following parameters are fixed: the ramp time $t_1= 0.02\text{s}$, the Mooney-Rivlin parameter $\alpha_{\text{MR}} = 1$, the elastic moduli $E_l = 75$, $\mu_l = 5$, $\mu_t= 1$, the long-term moduli $E_{l\infty}= 0.5$, $\mu_{l\infty} = 0.8$, $\mu_{t\infty} = 0.9$, and the characteristic times $\tau_{\mathcal{R}}= 1$, $\tau_6 = 1$ and $\tau_5 = 1$.}\label{fig:uni-axial}
\end{figure}

Let us compare the predictions of the two models under \textit{uni-axial elongation}. In this case, the fibres are stretched while the material is deformed, therefore $\hat{\m}\neq\m$ (see Table \ref{tab:deform-vectors}). By assuming the lateral surfaces to be free of traction, the only non-zero component of the Cauchy stress is $T_{33}(t)$, which can be calculated from \eqref{mQLV-TI-incompr} by using the corresponding deformation gradient in Table \ref{tab:deform-vectors} (see the Appendix for the detailed derivation):
\begin{equation}\label{uniaxial-m(t)}
\begin{split}
T_{33}(t)&=T_{33}^{\el}(t)+\dfrac{\Lambda^2(t)}{2}\int_0^t \left(R'(t-\tau ) \left(3 \Lambda^2 (\tau)-1\right)+2 R(t-\tau ) \left(1-6 \Lambda^2(\tau)\right) \dfrac{\Lambda'(\tau)}{\Lambda (\tau)}\right)  T_{33}^{\el}(\tau) \dt\\
&-\dfrac{1}{2\Lambda(t)}\int_0^t\Bigg(R_5'(t-\tau ) \Big(\Lambda^2(\tau)-1\Big)-2 R_5(t-\tau) \Lambda(\tau) \Lambda'(\tau)  \Bigg) \Lambda (\tau) T_{33}^{\el}(\tau)\dt\\
&+\dfrac{\Lambda^2(t)}{2}\int_0^t \Bigg(R_5'(t-\tau ) \Big(\Lambda^2(\tau)-1\Big)-4 R_5(t-\tau ) \Lambda (\tau) \Lambda'(\tau)\Bigg) \dfrac{\Lambda^2(\tau)-1}{ \Lambda^2(\tau)} T_{33}^{\el}(\tau)\dt\\
&+2\Lambda^2(t)\int_0^t \left(R_6'(t-\tau ) \left(1-\Lambda^2 (\tau)\right)+2R_6(t-\tau ) \left(2 \Lambda^2(\tau)-1\right) \dfrac{\Lambda'(\tau)}{\Lambda (\tau)}\right)T_{33}^{\el}(\tau) \dt.
\end{split}
\end{equation}
We recall below the expression for the stress $T_{33}$ derived in \cite{balbi2018modified} for the MQLV model under uni-axial extension, where the bases are written with respect to the normalised vector $\hat{\m}$:
\begin{equation}\label{T33hatm}
T_{33}(t)=T_{33}^{\el}(t)+\Lambda(t)^2\int_0^t\dfrac{\mathcal{R}'(t-\tau)}{E_l}\dfrac{T_{33}(\tau)}{\Lambda(\tau)}\dt.
\end{equation}
In Figure \ref{fig:uni-axial}, 
we plot the relaxation curves for uni-axial extension at different levels of stretch calculated using the MQLV model via Eqs. \eqref{uniaxial-m(t)} (left) and \eqref{T33hatm} (right), where the bases are written with respect to the vectors $\m(t)$ and $\hat{\m}(t)$, respectively.
Figure \ref{fig:uni-axial} shows that only the model written with respect to the deformed vector $\m(t)$ predicts strain-dependent relaxation. This phenomenon follows from the fact that the fibre vector $\m(t)$ accounts for the stretching of the fibres during the deformation. The more the fibres are stretched, the more the material relaxes after the initial deformation. However, the vector $\hat{\m}$ only accounts for changes in the fibres' orientation (not for changes in the magnitude of the stretching). In the small deformation limit (see the curves for $\Lambda=0.05$ in Figure \ref{fig:uni-axial}), the two versions of the MQLV model predict the same results, as expected. Indeed, $\hat{\m}\approx\m\approx\textbf{M}$.

\newpage
The other deformation that can accommodate non-linear features is \textit{perpendicular shear}.  Under this deformation, by assuming the lateral sides of the rectangular block are free of traction, the only non-zero stress components are the two normal stresses, i.e. $T_{11}(t)$ and $T_{33}(t)$, and the shear stress $T_{13}(t)$. In particular, we look at the expressions for the normal stress $T_{33}(t)$ and the shear stress $T_{13}(t)$. These are given by:
\begin{equation}\label{perp-t33}
\begin{split}
T_{33}(t)&=T_{33}^{\el}(t)+\int_0^t \mathcal{R}'(t-\tau ) \Bigg(\dfrac{1}{2} \Big(3 \kappa_3^2 (\tau)-1\Big) T_{11}^{\el}(\tau)+3 \kappa_3 (\tau) T_{13}^{\el}(\tau)+T_{33}^{\el}(\tau)\Bigg)\dt\\
&-3\int_0^t \mathcal{R}(t-\tau ) \kappa_3'(\tau) \Bigg(\kappa_3 (\tau) T_{11}^{\el}(\tau)+T_{13}^{\el}(\tau)\Bigg) \dt\\
& +\int_0^t R_5'(t-\tau ) \left(\dfrac{1}{2}\Big(3\kappa_3^2 (\tau)-1\Big) T_{11}^{\el}(\tau)-\kappa_3 (\tau) T_{13}^{\el}(\tau)\right)\dt\\
&-\int_0^tR_5(t-\tau )  \kappa_3'(\tau) \Bigg(3\kappa_3 (\tau) T_{11}^{\el}(\tau)+T_{13}^{\el}(\tau)\Bigg)\dt \\
&-2\int_0^t R_6'(t-\tau ) \kappa_3 (\tau) \Bigg(\kappa_3 (\tau) T_{11}^{\el}(\tau)+T_{13}^{\el}(\tau)\Bigg)\dt\\
&+2\int_0^t  R_6(t-\tau ) \kappa_3'(\tau) \Bigg(2 \kappa_3 (\tau) T_{11}^{\el}(\tau)+T_{13}^{\el}(\tau)\Bigg) \dt 
\end{split}
\end{equation}
and
\begin{equation}\label{perp-t13}
\begin{split}
T_{13}(t)=T_{13}^{\el}(t)&+\dfrac{\kappa_3 (t)}{2} \int_0^t \mathcal{R}'(t-\tau)\Bigg(\Big(3 \kappa_3^2(\tau)-1\Big) T_{11}^{\el}(\tau) +6 \kappa_3(\tau) T_{13}^{\el}(\tau) +2 T_{33}^{\el}(\tau) \Bigg)\dt\\
&-\dfrac{1}{2} \int_0^t\mathcal{R}(t-\tau) \kappa_3'(\tau)\Bigg(T_{11}^{\el}(\tau)  \Big(3 \kappa_3\big(\tau) (\kappa_3(\tau)+2 \kappa_3(t)\big)-1\Big)\\
&\hspace{10em}+6 T_{13}^{\el}(\tau) \Big(\kappa_3(\tau)+\kappa_3(t)\Big)+2 T_{33}^{\el}(\tau) \Bigg)\dt\\
&-\dfrac{1}{2} \int_0^tR_5'(t-\tau)\kappa_3(\tau)   \left(\kappa_3^2(\tau)-2 \kappa_3(\tau) \kappa_3 (t)+1\right)T_{11}^{\el}(\tau)\dt\\
&+\dfrac{1}{2} \int_0^tR_5(t-\tau)\kappa_3'(\tau)   \left(\kappa_3^2 (\tau)-4 \kappa_3 (\tau) \kappa_3 (t)+1\right)T_{11}^{\el}(\tau)\dt\\
& -\int_0^tR_6'(t-\tau)\Bigg(\kappa_3(\tau) \Big(\big(2 \kappa_3(\tau) \kappa_3(t)-1\big)T_{11}^{\el}(\tau) +T_{33}^{\el}(\tau)\Big)\\
&\hspace{7em}+\Big(\kappa_3(\tau)\big(\kappa_3(\tau)+2 \kappa_3(t)\big)-1\Big)T_{13}^{\el}(\tau)\Bigg)\dt\\
&+\int_0^tR_6(t-\tau)\kappa_3'(\tau ) \Bigg(\Big(2 \kappa_3 (\tau) \big(\kappa_3 (\tau)+2 \kappa_3 (t)\big)-1\Big)T_{11}^{\el}(\tau)\\
&\hspace{9.5em}+2  \big(2 \kappa_3 (\tau)+\kappa_3 (t)\big)T_{13}^{\el}(\tau)+T_{33}^{\el}(\tau)\Bigg)\dt,
\end{split}
\end{equation}
respectively.
\begin{figure}[t!]
\centering
\includegraphics[scale=0.36]{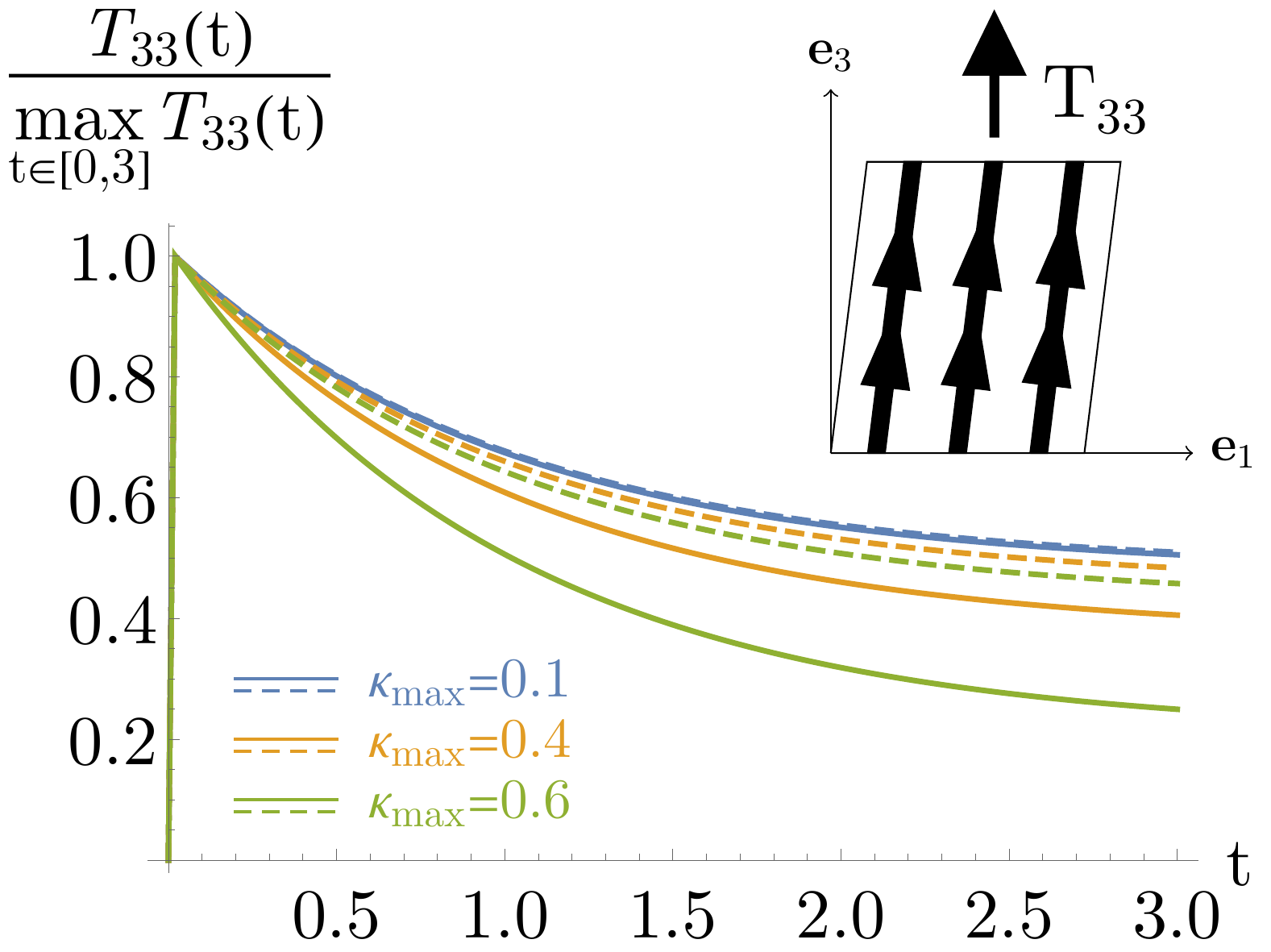}\hfill
\includegraphics[scale=0.36]{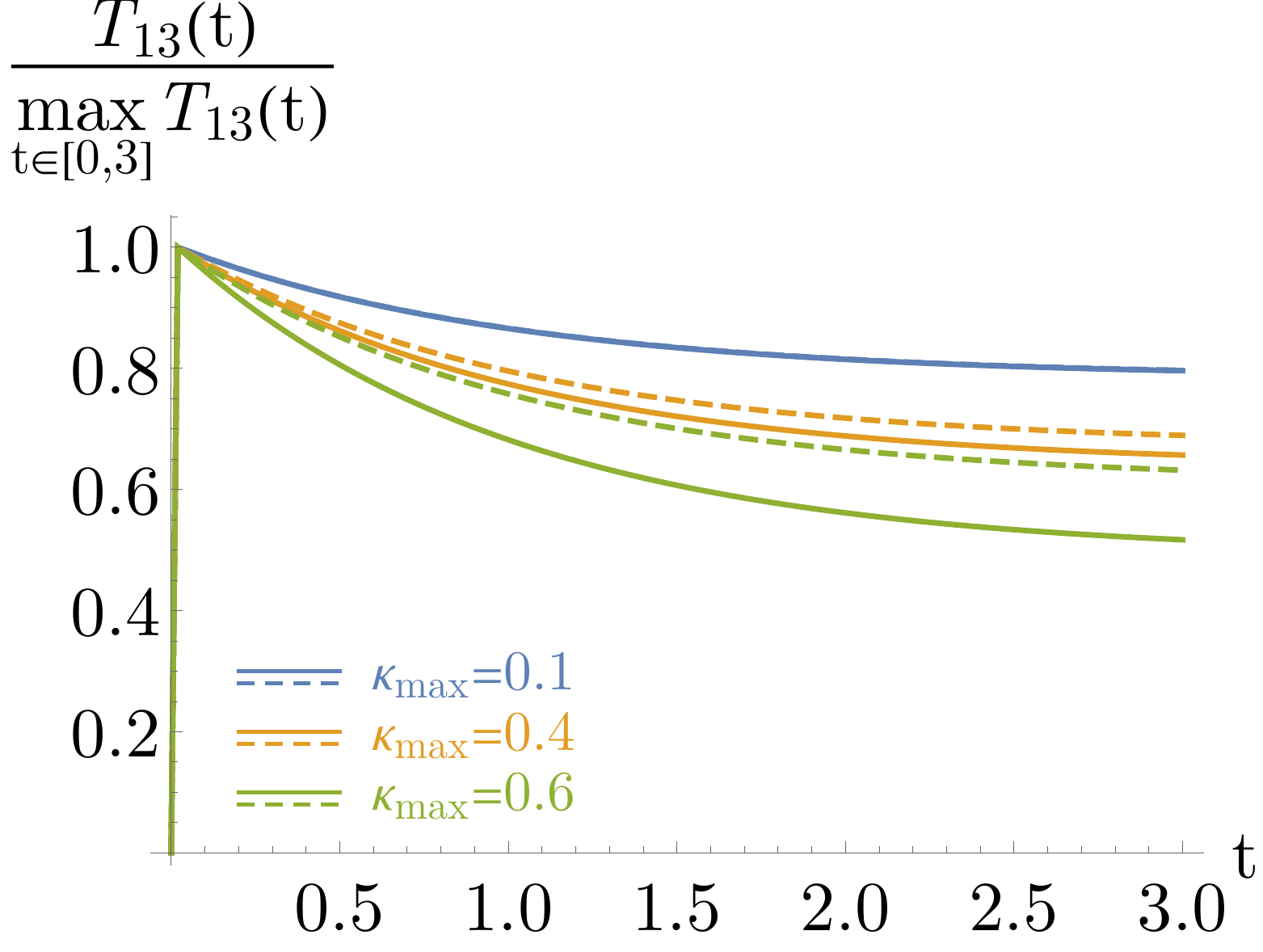}\caption{\textbf{Perpendicular shear}. Normalised stress relaxation curves for the normal (left) and shear (right) stresses $T_{33}(t)$ and $T_{13}(t)$, respectively, at different level of shear $\kappa_{\text{max}}=\{0.1,0.4,0.6\}$. The solid curves are plotted by using Eqs. \eqref{perp-t33} and \eqref{perp-t13},which are obtained from the version of the MQLV model where the bases are written with respect to the vector $\m$. The dashed curves are plotted by using Eqs. \eqref{T33} and \eqref{T13} with the bases written with respect to the vector $\hat{\m}$. The amount of shear $\kappa_3(t)$ is in the form of a ramp function.The following parameters are fixed: the ramp time $t_1= 0.02\text{s}$, the Mooney-Rivlin parameter $\alpha_{\text{MR}} = 1$, the elastic moduli $E_l = 75$, $\mu_l = 5$, $\mu_t= 1$, the long-term moduli $E_{l\infty}= 0.5$, $\mu_{l\infty} = 0.8$, $\mu_{t\infty} = 0.9$, and the characteristic times $\tau_{\mathcal{R}}= 1$, $\tau_6 = 1$ and $\tau_5 = 1$.}\label{fig:perp}
\end{figure}
We skip here the mathematical details associated with deriving \eqref{perp-t33} and \eqref{perp-t13}, which can be found in the Appendix. The normal stress $T_{33}(t)$ is the stress component perpendicular to the direction of shear and is therefore associated with the Poynting effect. We recall that the Poynting effect is a non-linear phenomenon, whereby the sheared material tends to either expand or contract in the direction perpendicular to the direction of shear. If the material tends to expand, the normal stress $T_{33}(t)$ is negative and a compressive force is required in order to maintain the deformation. Vice-versa, if the material tends to compress vertically, then $T_{33}(t)$ is positive and a tensile force is required to shear the material. In perpendicular shear, the fibres are stretched and rotate. Since the fibres are much stiffer than the isotropic matrix, they generate a vertical compression, which in turn results in a negative Poynting effect \cite{destrade2015dominant}. In Figure \ref{fig:perp}, we plot the relaxation curves for the normal (left) and shear (right) stresses for the MQLV model written with respect to $\m$ (solid lines) and $\hat{\m}$ (dashed lines). First, we note that in perpendicular shear both versions of the MQLV model predict strain-dependent relaxation. The higher the level of shear, the more significant the relaxation. Moreover, the MQLV model written with respect to the vector $\m$ predicts a more significant relaxation than the MQLV model written with respect to $\hat{\m}$. This is a direct consequence of the fact that in perpendicular shear, the fibres stretch and rotate in the $\textbf{e}_1$-$\textbf{e}_3$ plane. However, the vector $\m$ accounts for both the stretching and rotation of the fibres, whereas the vector $\hat{\m}$ only accounts for the rotation of the fibres and is unaffected by the stretching. 

\section{Conclusion}
The theory of quasi-linear viscoelasticity has been criticised in the past for its limitations, in particular for failing to capture the non-linearities associated with strain-dependent relaxation and stress-dependent creep. In this paper, we have investigated a modified formulation of the QLV theory for TI materials that accounts for strain-dependent relaxation. The key feature of our model is the linear decomposition of the tensorial relaxation function $\mathbb{G}$ into the sum of scalar components associated with a set of fourth-order tensorial bases. The scalar components are the relaxation functions inherited from the linear viscoelastic theory. We have shown that the set of bases must satisfy the property of additive symmetry in order for the constitutive equation to recover the elastic limit. We have proposed a robust method to identify such a set of bases both for isotropic and transversely isotropic materials. For fibre-reinforced TI materials, the bases naturally depend on the deformation through the fibre vector. We have discussed two alternative formulations. The first uses the bases with respect to the normalised deformed fibre vector $\hat{\m}(t)$. In the second version, we write the bases with respect to the deformed vector $\m(t)$. Finally, we have shown that the two models are able to capture strain-dependent relaxation, opening the way towards a fully non-linear viscoelastic theory.

\section{Appendix}
\subsection{Uni-axial elongation of a TI material}
In uni-axial elongation, we assume that the material is stretched by an amount $\Lambda(t)$ along $\textbf{e}_3$:
\begin{equation}
x_1(t)=\nicefrac{1}{\sqrt{\Lambda(t)}}X_1,\qquad
x_2(t)=\nicefrac{1}{\sqrt{\Lambda(t)}}X_2,\qquad
x_3(t)=\Lambda(t)X_3.
\end{equation}
From the constitutive equations \eqref{mQLV-TI-incompr}, it follows that the non-zero Cauchy stress components are:
\begin{equation}
T_{11}(t)=T_{22}(t)=-p(t)+\dfrac{1}{\Lambda(t)}\left(\PPi_{\Tv11}(t)+\int_0^t \Big(R_5'(t-\tau )\PPi_{\Tv11}(\tau)-R_5(t-\tau) \Om_{\Tv11}(\tau)\Big) \dt\right),
\end{equation}
and
\begin{equation}\label{T33uniax-prep}
\begin{split}
T_{33}(t)=-p(t)&+\Lambda (t)^2 \left(\PPi_{\An33}(t)+\PPi_{\Lo33}(t)+\PPi_{\Tv33}(t)\right)\\
&+\Lambda (t)^2 \left(\int_0^t \Big(\mathcal{R}'(t-\tau ) \PPi_{\Lo33}(\tau)-\mathcal{R}(t-\tau) \Om_{\Lo33}(\tau)\Big) \dt\right.\\
&+\int_0^t \Big(R_5'(t-\tau ) \PPi_{\Tv33}(\tau)-R_5(t-\tau ) \Om_{\Tv33}(\tau)\Big) \dt\\
& \left.+\int_0^t \Big(R_6'(t-\tau ) \PPi_{\An33}(\tau)-R_6(t-\tau ) \Om_{\An33}(\tau)\Big) \dt\right).
\end{split}
\end{equation}
By assuming no lateral traction on the surfaces with outer unit normals $\textbf{n}_1=\textbf{e}_1$ and $\textbf{n}_2=\textbf{e}_2$, it follows that $\textbf{T}\textbf{e}_1=T_{11}=0$. From this condition, we can obtain the Lagrange multiplier p(t) as follows:
\begin{equation}\label{p-uniax}
p(t)=\dfrac{1}{\Lambda(t)}\left(\PPi_{\Tv11}(t)+\int_0^t \Big(R_5'(t-\tau )\PPi_{\Tv11}(\tau)-R_5(t-\tau) \Om_{\Tv11}(\tau)\Big) \dt\right).
\end{equation}
Now, by substituting Eq. \eqref{p-uniax} into Eq. \eqref{T33uniax-prep} and by calculating the terms $\PPi_i^{\el}$ and $\Om_i^{\el}$ ($i=\{\Lo,\Tv,\An\}$) from Eq. \eqref{Pi-Om-TI} and the bases in Eq. \eqref{symm-add-Hill}, we get the expression for the normal stress $T_{33}$ in Eq. \eqref{uniaxial-m(t)}.

\subsection{Perpendicular shear of a TI material}\label{app-perp}
In perpendicular shear, a TI incompressible material deforms according to the following equations:
\begin{equation}
x_1(t)=X_1+\kappa_3(t)X_3,\qquad
x_2(t)=X_2,\qquad
x_3(t)=X_3,
\end{equation}
where $\kappa_3(t)$ is the amount of shear in the $\textbf{e}_1$-$\textbf{e}_3$ plane. The deformation gradient is given in Table \ref{tab:deform-vectors}. Neglecting acceleration, the governing equations $\text{div}\textbf{T}(t)=\boldsymbol{0}$ are identically satisfied, where $\text{div}$ is the divergence operator, calculated with respect to the deformed coordinates $x_i$.
It follows that the normal and shear stresses can be obtained directly from Eq. \eqref{mQLV-TI-incompr} and are given by:
\begin{equation}\label{T33}
\begin{split}
T_{33}(t)&=-p(t)+\PPi_{\Lo33}^{\el}(t)+\PPi_{\Tv33}^{\el}(t)+\PPi_{\An33}^{\el}(t)\\
&\int_0^t \Bigg(\mathcal{R}'(t-\tau )\PPi_{\Lo33}^{\el}(\tau ) -\mathcal{R}(t-\tau )\Om_{\Lo33}^{\el}(\tau)\Bigg) \dt\\
&+\int_0^t\Bigg(R_5'(t-\tau ) \PPi_{\Tv33}^{\el}(\tau)
- R_5(t-\tau)\Om_{\Tv33}^{\el}(\tau)\Bigg)\dt \\
&+\int_0^t \Bigg(R_6'(t-\tau)\PPi_{\An33}^{\el}(\tau) -R_6(t-\tau )\Om_{\An33}^{\el}(\tau)\Bigg)\dt,
\end{split}
\end{equation}
\begin{equation}\label{T13}
\begin{split}
T_{13}(t)&=\PPi_{\Tv13}^{\el}(t) +\PPi_{\An13}^{\el}(t)+\kappa_3(t)\Big(\PPi_{\Lo33}^{\el}(t)+\PPi_{\Tv33}(t)+\PPi_{\An33}^{\el}(t)\Big)\\
&+\int_0^t \Bigg(R_5'(t-\tau )\PPi_{\Tv13}^{\el}(\tau)-R_5(t-\tau )\Om_{\Tv13}^{\el}(\tau)\Bigg)\dt\\
&+\int_0^t \Bigg(R_6'(t-\tau)\PPi_{\An13}^{\el}(\tau)-R_6(t-\tau )\Om_{\An13}^{\el}(\tau)\Bigg)\dt \\
& +\kappa_3(t)\Bigg(\int_0^t \Big(\mathcal{R}'(t-\tau)\PPi_{\Lo33}^{\el}(\tau )-\mathcal{R}(t-\tau )\Om_{\Lo33}^{\el}(\tau)\Big) \dt\\
& +\int_0^t \Big(R_5'(t-\tau ) \PPi_{\Tv33}^{\el}(\tau) - R_5(t-\tau )\Om_{\Tv33}^{\el}(\tau) \Big)\dt \\
&+\int_0^t \Big(R_6'(t-\tau ) \PPi_{\An33}(\tau) -R_6(t-\tau )\Om_{\An33}(\tau) \Big)\dt\Bigg).
\end{split}
\end{equation}
The Lagrange multiplier $p(t)$ can be obtained by imposing the condition of zero lateral traction on the surface with normal $\textbf{n}_2=(0,1,0)^{\tra}$ in the current configuration. The boundary condition can be written as $\textbf{T}(t)\textbf{n}_2=T_{22}(t)=0$, $\forall t$ and gives the following equation for $p(t)$:
\begin{equation}\label{p-perpe}
p(t)=\int_0^t R_5'(t-\tau )\PPi_{\Tv22}(\tau) \dt-\int_0^tR_5(t-\tau )\Om_{\Tv22}^{\el}(\tau) \dt +\PPi_{\Tv22}^{\el}(t).
\end{equation}
The terms $\PPi_i^{\el}$ and $\Om_i^{\el}$ ($i=\{\Lo,\Tv,\An\}$) can be calculated by using Eq. \eqref{Pi-Om-TI} and the bases in Eq. \eqref{symm-add-Hill}. 
\bibliographystyle{vancouver}
\bibliography{biblio-bases.bib}

\end{document}